\documentstyle[12pt, twoside, epsf, here, psfig]{article}
\newcounter{tony}
\newcommand{\bb}{\mbox{\boldmath{$b$}}}

\newcommand{\bn}{\mbox{\boldmath{$n$}}}

\newcommand{\bu}{\mbox{\boldmath{$u$}}}
\newcommand{\bv}{\mbox{\boldmath{$v$}}}
\newcommand{\bw}{\mbox{\boldmath{$w$}}}
\newcommand{\bx}{\mbox{\boldmath{$x$}}}

\newcommand{\bA}{\mbox{\boldmath{$A$}}}
\newcommand{\bB}{\mbox{\boldmath{$B$}}}

\newcommand{\bD}{\mbox{\boldmath{$D$}}}

\newcommand{\bF}{\mbox{\boldmath{$F$}}}

\newcommand{\bI}{\mbox{\boldmath{$I$}}}

\newcommand{\bN}{\mbox{\boldmath{$N$}}}

\newcommand{\bP}{\mbox{\boldmath{$P$}}}

\newcommand{\blambda}{\mbox{\boldmath{$\lambda$}}}

\newcommand{\bsigma}{\mbox{\boldmath{$\sigma$}}}

\newcommand{\beq}{\begin{equation}}
\newcommand{\eeq}[1]{\label{eq:#1}\end{equation}}

\newcommand{\eqref}[1]{(\ref{eq:#1})}

      \newcommand{\beqn}{\begin{equation}}
      \newcommand{\eeqn}{\end{equation}}
      \newcommand{\beqna}{\begin{eqnarray}}
      \newcommand{\eeqna}{\end{eqnarray}}

\newtheorem{lemma}{\sc Lemma}
\newtheorem{theorem}{\sc Theorem}
\newtheorem{inequality}{\sc Inequality}
\newtheorem{assumption}{\sc Assumption}

\title{The Energetic Implications of Using Deforming Reference
Descriptions to Simulate the Motion of Incompressible, Newtonian
Fluids}

\author{S. J. Childs \\ \\ {\small\em Department of Pure and Applied
Mathematics, Rhodes University, Grahamstown,} \\ {\small\em 6140, South
Africa}}

\renewcommand{\thefootnote}{\fnsymbol{footnote}}

\date{}       

\begin{document}

\maketitle

\renewcommand{\thefootnote}{\arabic{footnote}}

\begin{abstract}
\noindent {\em In this work the issue of whether key energetic
properties (nonlinear, exponential--type dissipation in the absence of
forcing and long--term stability under conditions of time dependent
loading) are automatically inherited by deforming reference
descriptions is resolved. These properties are intrinsic to real flows
and the conventional Navier--Stokes equations. A completely general
reference description of an incompressible, Newtonian fluid, which
reconciles the differences between opposing schools of thought in the
literature is derived for the purposes of this investigation. \\

\noindent The work subsequently focusses on establishing a class of
time discretisations which inherit these self--same energetic
properties, irrespective of the time increment employed. The findings
of this analysis have profound consequences for the use of certain
classes of finite difference schemes in the context of deforming
references. It is significant that many algorithms presently in use do
not automatically inherit the fundamental qualitative features of the
dynamics. An ``updated'' approach as a means of avoiding ever
burgeoning deformation gradients and a still further simplified
implementation are further topics explored.} \end{abstract}


\section{Introduction}

Descriptions of fluid motion are conventionally based on the principles
of conservation of mass and linear momentum. One might hope that all
such descriptions would accordingly exhibit key energetic properties
(nonlinear, exponential--type dissipation in the absence of forcing and
long--term stability under conditions of time dependent loading)
consistant with the principle of energy conservation. These properties
are intrinsic to real flows and the conventional, Eulerian
Navier--Stokes equations.

A completely general reference description of an incompressible,
Newtonian fluid, which reconciles the differences between the
so--called arbitrary Lagrangian Eulerian (A.L.E.) formulation of {\sc
Hughes, Liu} and {\sc Zimmerman} \cite{h:1} (deformation gradients
absent) and that of {\sc Soulaimani, Fortin, Dhatt} and {\sc Ouellet}
\cite{f:1} (deformation gradients present, but use is problematic), is
derived for the purposes of this investigation. The implications of the
resulting description are investigated in the context of energy
conservation in a similar, but broader, approach to that taken by
others (eg. {\sc Simo} and {\sc Armero} \cite{s:1}) for the
conventional, Eulerian Navier--Stokes equations.

The work subsequently focusses on establishing a class of time
discretisations which inherit these self--same energetic properties
irrespective of the time increment employed. The findings of this
analysis have profound consequences for the use of certain classes of
difference schemes in the context of deforming references. It is
significant that many algorithms presently in use do not automatically
inherit the fundamental qualitative features of the dynamics. An
``updated'' approach as a means of avoiding ever burgeoning deformation
gradients which arise from the accumulated step--wise deformation of
meshes and a still further simplified implementation are further topics
explored.

The main conclusions of this work are based on a new inequality and a
number of lemmas. These lemmas are mainly concerned with the new
convective term. The new inequality is used in place of where the
Poincar\'{e}--Friedrichs inequality might otherwise have limited the
analysis. This analysis is extended in that non--zero boundaries,
so--called free boundaries and time--dependent loads are considered.

\section{A Completely General Reference} \label{130}

The implementation of most numerical time integration schemes would be
problematic were a conventional Eulerian\footnotemark[1] description of
fluid motion to be used in instances involving deforming domains. The
reason is that most numerical time integration schemes require
successive function evaluation at fixed spatial locations. On the other
hand meshes rapidly snarl when purely Lagrangian\footnotemark[2]
descriptions are used. \footnotetext[1]{{\sc Eulerian} or {\sc spatial}
descriptions are in terms of fields defined over the current
configuration.} \footnotetext[2]{{\sc Lagrangian} or {\sc material}
descriptions are made in terms of fields defined over a reference
(material) configuration.}

Eulerian and Lagrangian references are just two, specific examples of
an unlimited number of configurations over which to define fields used
to describe the dynamics of deforming continua. They are both special
cases of a more general reference description, a description in which
the referential configuration is deformed at will. A deforming finite
element mesh would be a good example of just such a deforming reference
in practice. The transformation to the completely general reference
involves coordinates where used as spatial variables only and the
resultant description is therefore inertial in the same way as
Lagrangian descriptions are.

\subsection{Notation}

Consider a material body which occupies a domain $\Omega$ at time $t$.
The material domain, $\Omega_0$, is that corresponding to time $t =
t_0$ (the reference time, $t_0$, is conventionally, but not always,
zero). A third configuration, ${\tilde \Omega}$, which is chosen
arbitrarily is also defined for the purposes of this work. The three
domains are related in the sense that points in one domain may be
obtained as one--to--one invertible maps from points in another.
\begin{figure}[h]
\begin{center}
\mbox{\epsfbox{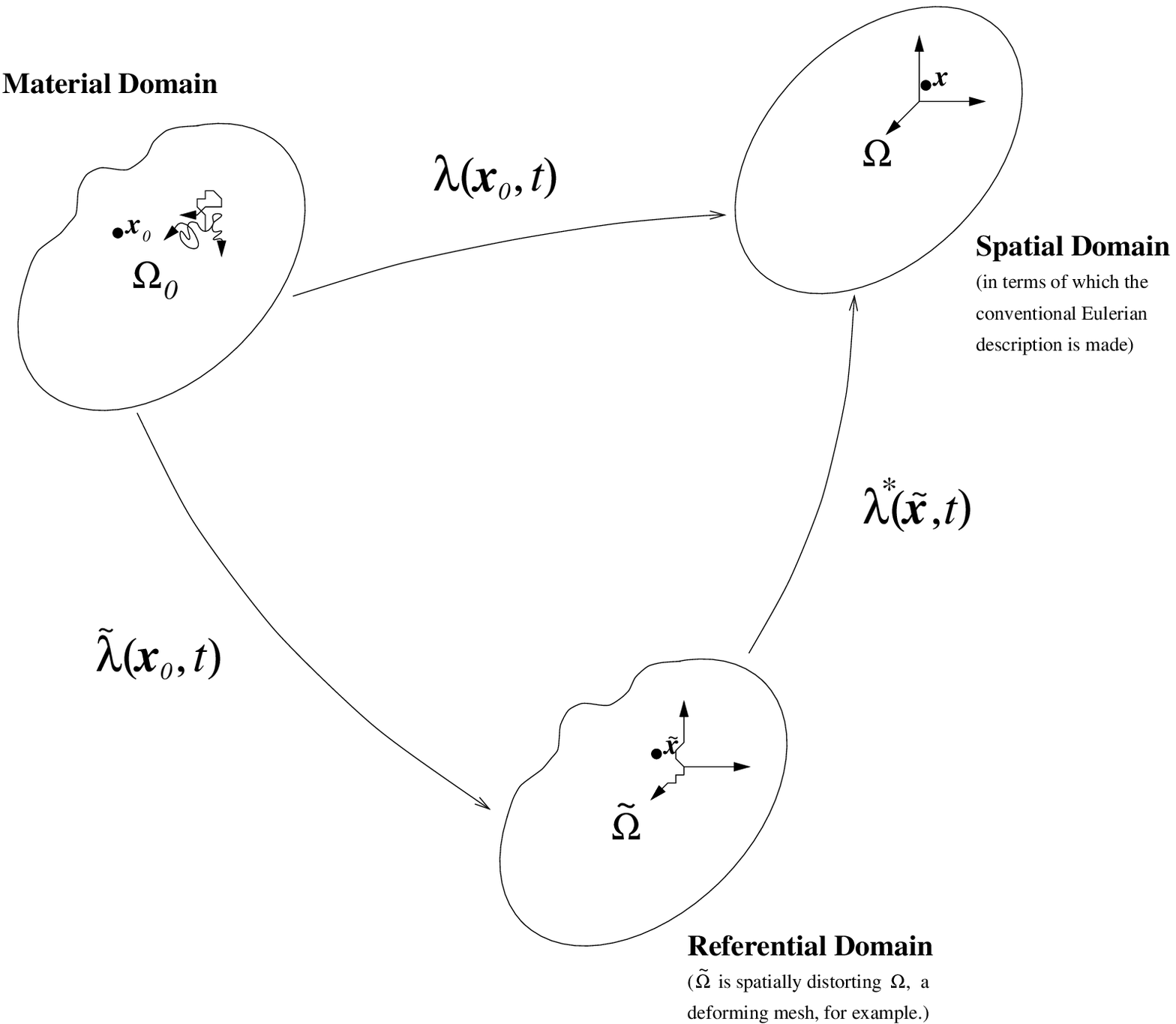}}
\end{center}
\caption{Schematic Diagram of Domains and Mappings Used in a
Completely General Reference Description} \label{91}
\end{figure}

For any general function $f({\bx},t)$, a function, $\tilde
f(\tilde{\bx},t) \ \equiv \ f({\blambda}^{\ast}(\tilde{\bx},t),t)$, can
be defined in terms of the domains and one--to--one, invertible
mappings illustrated in Figure \ref{91}. Similarly, $f_0({\bx}_0 ,t)
\ \equiv \ f({\blambda}({\bx}_0 ,t),t)$ can be defined. This notation
can be generalised for the component--wise definition of higher order
tensors. The key to understanding much of this work lies possibly in
adopting a component-wise defined notation.

In contrast to the function notation just established, the definition
of the operators ${\tilde \nabla}$ and $\widetilde {\mbox{div}}$ is not
based on $\nabla$ and $\mbox{div}$. They are instead the referential
counterparts, that is
\[ 
{\tilde \nabla} = \frac{\partial}{\partial {\tilde {\bx}}}
\hspace{10mm} \mbox{and} \hspace{10mm} {\widetilde {\mbox{div}}} =
\frac{\partial}{\partial {\tilde x}_1} + \frac{\partial}{\partial
{\tilde x}_2} + \frac{\partial}{\partial {\tilde x}_3}.
\] 
The notation ${\bA}:{\bB}$ is used to denote the matrix inner product
$A_{ij}B_{ij}$ throughout this work, $\left< \ \cdot \ , \ \cdot
\ \right>_{L^2( \ \cdot \ )}$ denotes the $L^2$ inner product and
$\left|\left| \ \cdot \ \right|\right|_{L^2( \ \cdot \ )}$ the $L^2$
norm.

\subsection{Some General Results for Functions Defined on the Three
Domains}

Three important results are necessary for the derivation of the
completely general reference description and these are presented
below.

\subsubsection*{The Material Derivative in Terms of a Completely
General Reference}

The material derivative of any vector field ${\tilde {\bv}}$ in terms
of a completely general, reference is
\begin{eqnarray} \label{25} 
\frac{\partial \tilde{\bv}}{\partial t} + {{\tilde \nabla} \tilde
{\bv}} \left[ {\tilde {\bF}}^{-1} (\tilde{\bv} - \tilde{\bv}^{ref})
\right].
\end{eqnarray}      
where $\tilde{\bv}^{ref}$ is the velocity of the reference deformation,
and $\tilde {\bF}$ is the deformation gradient given by
\[
{\tilde {\bF}}({\tilde {\bx}}) = \frac{\partial {\blambda}^*}{\partial
{\tilde {\bx}}}.
\]
This result (taken from {\sc Hughes, Liu} and {\sc Zimmerman}
\cite{h:1}) is obtained by recalling that the material derivative
(total derivative) is the derivative with respect to time in the
material configuration. Thus
\begin{eqnarray} \label{26}
\frac{D\tilde{v}_i}{Dt} & = & \frac{\partial}{\partial t} \{ {\tilde
v}_i ( {\tilde {{\blambda}}} ({\bx} _0,t) ,t)  \} \nonumber \\
& = & \frac{\partial \tilde{v}_i}{\partial t} + \frac{\partial
\tilde{v}_i}{\partial \tilde{x}_j}\frac{\partial
\tilde{\lambda}_j}{\partial t} \ .
\end{eqnarray}
A more practical expression is needed for $\displaystyle \frac
{\partial\tilde{\lambda_j}} {\partial t}$ (the velocity as perceived in
the distorting reference). This can be obtained by considering
\[
\lambda_k({\bx} _0,t) = \lambda^{\ast}_k(\tilde {{\blambda}} ({\bx}
_0,t),t) \hspace{10mm} \mbox{(see Figure \ref{91} on page
\pageref{91})}
\] 
so that
\[ 
{\left.\frac{ \partial \lambda_k }{\partial t} \right|} _{ {\bx} _0
\ fixed} = {\left.\frac{ \partial \lambda^{\ast}_k }{\partial t}
\right|}_{ {\tilde{\bx} } \ fixed} + \frac{ \partial \lambda^{\ast}_k
}{\partial \tilde{x}_j}\frac{\partial \tilde{\lambda}_j}{\partial t}
\]
or
\[ 
\frac{\partial \tilde{\lambda}_j}{\partial t} = \frac{\partial
\tilde{x}_j}{\partial x_k} \left( {\left. \frac{ \partial \lambda_k
}{\partial t} \right|}_{ {\bx} _0 \ fixed} - {\left. \frac{ \partial
\lambda^{\ast}_k }{\partial t} \right|}_{ {\tilde {\bx} } \ fixed}
\right).
\] 
Substituting this expression into equation (\ref{26}), the desired,
suitably practicable result is obtained.

\subsubsection*{An Element of Area in Terms of a Distorting Reference} 

The second important result can be recalled from general continuum
mechanics. Consider an element of area, size $dA$, with an outward unit
normal ${\bn}$. Then
\begin{eqnarray} \label{27}
{\bn} dA = {\tilde{\bF}}^{-t}\tilde{\bN} {\tilde J}
d\tilde{A} 
\end{eqnarray} 
where $d{\tilde A}$ and $\tilde{\bN}$ denote the respective analogous
size and outward unit normal of this element of area in the referential
configuration (capital ``n'' so as to remain consistent with the
notation, since ${\tilde N}_i \neq n_i$ in this case) and ${\tilde J}
= \det{\tilde {\bF}}$. This result is demonstrated in most popular
textbooks on continuum mechanics (eg. {\sc Lai}, {\sc Rubin} and {\sc
Krempl} \cite{lrk:1}).

\subsubsection*{The Kinematic Result ${\dot {\cal J}}_0 = {\cal J}_0
\mathop{\rm div}{\bv}$} \label{5}

The material derivative of the Jacobian ${\cal J}_0$ is given by the
relation
\[
{\dot {\cal J}}_0 = {\cal J}_0 \mathop{\rm div}{\bv}
\]
where ${\cal J}_0$ is defined as follows,
\[
{\cal J}_0 \equiv \det \left\{ \frac{\partial {\blambda}}{\partial
{\bx}_0} \right\}.
\]
This result is demonstrated in most popular textbooks on continuum
mechanics (eg. {\sc Lai}, {\sc Rubin} and {\sc Krempl} \cite{lrk:1}).

\subsection{Derivation of the Completely General Equation}

One way in which to derive a completely general reference
description of an incompressible, Newtonian fluid is to start with
the balance laws in global (integral) form, and to make the necessary
substitutions in these integrals. The desired numerical
implementation (similar to the conventional Navier--Stokes one which
has been thoroughly investigated and found to be stable) is then
obtained.

\subsubsection*{Conservation of Mass}

Let $\Omega(t)$ be an arbitrary sub--volume of material. The principle
of conservation of mass states that
\begin{eqnarray} \label{92}
\frac{d}{dt} \int_{\Omega(t)} \rho {d \Omega} &=& 0 \hspace{10mm}
\mbox{\it (rate of change of mass with time }= \ \mbox{\it 0)}
\nonumber \\ 
& & \nonumber \\
\frac{d}{dt} \int_{\Omega_0} {{\rho}_0} {\cal J}_0 {d \Omega_0} &=& 0
\hspace{10mm} \mbox{\it (reformulating in terms of the material}
\nonumber \\
& & \hspace{13mm} \mbox{\it configuration, } \Omega_0 \mbox{\it .)}
\nonumber \\
\int_{\Omega_0} \frac{\partial}{\partial t}  \left\{ {{\rho}_0} {\cal
J}_0 \right\} {d \Omega_0} &=& 0 \hspace{10mm} \mbox{\it (since limits
are not time dependent in} \nonumber \\
& & \hspace{13mm} \mbox{\it the material configuration.)} \\
\int_{\Omega_0} \left( \rho_0 {\dot {\cal J}}_0 \ + \ {\dot \rho}_0
{\cal J}_0 \right) d \Omega_0 &=& 0 \hspace{10mm} \mbox{\it (by the
chain rule)} \nonumber \\
& & \nonumber \\
\int_{\Omega(t)} \left( \dot{\rho} \ + \ {\rho \mathop{\rm div}{\bv}
} \right) {d \Omega} &=& 0 \hspace{10mm} \mbox{\it (using the
kinematic result } \dot{{\cal J}_0} = {\cal J}_0 \mbox{div}\, {\bv}
\mbox{\it )} \nonumber \\
& & \nonumber \\
\int_{{\tilde \Omega}(t)} \left( {\dot {\rho}} \ + \ {\rho}
\frac{\partial \tilde{v}_i}{\partial \tilde{x}_j}\frac{\partial
\tilde{x}_j}{\partial x_i} \right) {\tilde J} {d
{\tilde \Omega}} &=& 0 \hspace{10mm} \mbox{\it (reformulating in terms
of the distorting} \nonumber \\
& & \hspace{13mm} \mbox{\it referential configuration, }{\tilde \Omega}
(t) \mbox{\it .)} \nonumber \\
\Rightarrow \left( {\dot {\rho}} \ + \ {\rho} {{{\tilde \nabla} \tilde
{\bv} } : {\tilde {\bF} }^{-t}} \right) {\tilde J} &=&
0 \hspace{10mm} \mbox{\it (integrand must be zero since the volume}
\nonumber \\
& & \hspace{13mm} \mbox{\it was arbitrary.)} \nonumber 
\end{eqnarray} 
Thus, for a material of constant, non--zero density,
\[
{\tilde \nabla} {\tilde {\bv}} : {\tilde {\bF}}^{-t} = 0 \hspace{7mm}
\mbox{since} \hspace{7mm} {\tilde J} \neq 0
\hspace{12mm} \mbox{\it (mappings are one-to-one and invertible).}
\]
Notice also that equation (\ref{92}) implies 
\begin{eqnarray} \label{93}
\frac{\partial}{\partial t}  \left\{ {{\rho}_0} {\cal J}_0 \right\} &=&
0
\end{eqnarray}
since the volume was arbitrary and the integrand must therefore be
zero.

\subsubsection*{Conservation of Linear Momentum (and Mass)}

The principle of conservation of linear momentum for an arbitrary
volume of material $\Omega(t)$ with boundary $\Gamma(t)$ states that
\begin{eqnarray} \label{28} 
\frac{d}{dt} \int_{\Omega(t)} \rho {\bv} {d \Omega} = \int_{\Omega(t)}
\rho {\bb} {d \Omega} \ + \  \int_{\Gamma(t)} {\bsigma} {\bn} dA
\end{eqnarray}
where ${\rho}$ is density, ${\bb}$  is the body force per unit mass,
${\bsigma}$ is the stress, {\bn}  the outward unit normal to the
boundary and {\bv}  is the velocity. The term on the lefthand side can
be rewritten as follows:
\begin{eqnarray*}
{\frac{d}{dt}} \int_{\Omega(t)} \rho {\bv}  {d \Omega} &=&
\frac{d}{dt} \int_{\Omega_0} {{\rho}_0} {{\bv} _0} {\cal J}_0 {d
\Omega_0} \hspace{10mm} \mbox{\it (Reformulating in terms of the
material} \\
& & \hspace{42mm} \mbox{\it configuration, } \Omega_0\mbox{\it .)} \\ 
&=& \int_{\Omega_0} \frac{\partial}{\partial t}  \left\{ {{\rho}_0}
{{\bv} _0} {\cal J}_0  \right\} {d \Omega_0} \hspace{5mm} \mbox{\it (Since
limits are not time dependent in} \\
& & \hspace{41mm} \mbox{\it the material configuration.)} \\
&=&  \int_{\Omega_0} \left( \frac{\partial {\bv}_0}{\partial t}
{{\rho}_0} {\cal J}_0 \ + \ {{\bv}_0}\frac{\partial}{\partial t} \left\{
{{\rho}_0} {\cal J}_0 \right\} \right) {d \Omega_0} \\
& &  \\
& &  \\
&=& \int_{\Omega(t)} \rho \dot{\bv}  {d \Omega} \hspace{22mm} \mbox{\it
(The second term above is zero as a} \\
& & \hspace{41mm} \mbox{\it consequence of equation (\ref{93}).)} \\
&=& \int_{{\tilde \Omega}(t)} {\rho} {\dot{\tilde {\bv} }} {\tilde J}
{d {\tilde \Omega}}  \hspace{20mm} \mbox{\it (Reformulating in terms of
the dist--} \\
& & \hspace{42mm} \mbox{\it orting referential configuration, } \tilde
\Omega \mbox{\it .)} \\
&=& \int_{{\tilde \Omega}(t)} {\rho} \left( \frac{\partial
\tilde{\bv}}{\partial t} + { {{\tilde \nabla} \tilde {\bv} } } \left[
{\tilde {\bF}}^{-1} (\tilde{\bv}  - \tilde{\bv} ^{ref}) \right]
\right) {\tilde J} {d {\tilde \Omega}} \hspace{5mm}
\mbox{\it (Using result} \\
&& \hspace{79mm} \mbox{\it (\ref{25}) on page \pageref{25})}
\end{eqnarray*} 
where $\dot{\bv}$ denotes the material derivative of ${\bv} $. The
surface integral becomes
\begin{eqnarray*}
\begin{array}{rcll}
\displaystyle \int_{\Gamma(t)} {\bsigma} {\bn}  dA &=& \displaystyle
\int_{\tilde {\Gamma}(t)} {\tilde {\bsigma}} \tilde{\bF} ^{-t} {\tilde
{\bN}} {\tilde J} d{\tilde A} & \mbox{\it (Reformulating in terms of a
distorting} \\
& & & \mbox{\it \ reference using result (\ref{27}) on page
\pageref{27}.)} \\
&=& \displaystyle \int_{{\tilde \Omega}(t)} \mathop{\widetilde {\rm
div}}\, \{ {\tilde {\bsigma}} {\tilde{\bF} ^{-t}} {\tilde J} \} {d
{\tilde {\Omega}} } & \mbox{\it (By the divergence theorem).}
\end{array} 
\end{eqnarray*} 
Finally, the term involving body force becomes
\begin{eqnarray*}
\begin{array}{rcll}
\displaystyle \int_{\Omega(t)} \rho {\bb}  {d \Omega} &=& \displaystyle
\int_{{\tilde \Omega}(t)} {\rho} {\tilde {\bb} } {\tilde J} {d {\tilde
\Omega}} \hspace{18mm} & \mbox{\it (Reformulating in terms of a
distorting} \\
& & & \mbox{\it \ reference.).}   
\end{array}
\end{eqnarray*}
Substituting these expressions into (\ref{28}), remembering that the
volume used in the argument was arbitrary and that the entire integrand
must therefore be zero, the conservation principles of linear momentum
and mass may be written in primitive form as
\begin{eqnarray} \label{29}
{\rho} \left( \frac{\partial {\tilde {\bv} }}{\partial t} + {{{\tilde
\nabla} \tilde {\bv} } {\tilde {\bF} }^{-1} } ({\tilde {\bv} } -
{\tilde {\bv} }^{ref}) \right) {\tilde J} &=& {\rho} {\tilde {\bb} }
{\tilde J} + \mathop{\widetilde {\rm div}}{\tilde {\bP}}
\end{eqnarray}
and
\begin{eqnarray} \label{30}
{{\tilde \nabla} {\tilde {\bv}} : {\tilde {\bF} }^{-t} } \ = \ 0 
\end{eqnarray}
where $\tilde{\bP}$ is the Piola--Kirchoff stress tensor of the first
kind, $\tilde{\bP}  \ = \ \tilde{\bsigma} {\tilde {\bF}}^{-t} {\tilde
J}$. In terms of the constitutive relation, $\bsigma = - p {\bI} + 2
\mu {\bD}$, for a Newtonian fluid,
\[
{\tilde {\bP}} = \left( - p{\bI} + \mu \left[ {\tilde \nabla}{\tilde
{\bv}}{\tilde {\bF}}^{-1} + \left({\tilde \nabla}{\tilde {\bv}}{\tilde
{\bF}}^{-1}\right)^t \right] \right){\tilde {\bF}}^{-t} {\tilde J}
\hspace{8mm} \mbox{since} \hspace{8mm} {\tilde {\bD}} = \frac{1}{2}
\left( {\widetilde {\nabla {\bv}}} + \left( {\widetilde{\nabla {\bv}}}
\right)^t \right).
\]
The derivation of a variational formulation is along similar lines as
that for the Navier-Stokes equations (the purely Eulerian
description). For a fluid of constant density, the variational
formulation
\renewcommand{\thefootnote}{\fnsymbol{footnote}}
\begin{eqnarray} \label{33}
{\rho} \int_{ \tilde \Omega } {\tilde {\bw} } \cdot
\frac{\partial {\tilde {\bv} }}{\partial t} {\tilde J}
{d{\tilde \Omega}} \ + \ {\rho} \int_{ \tilde \Omega } {\tilde
{\bw}} \cdot {{\tilde \nabla} {\tilde {\bv} }} \left[{\tilde {\bF}
}^{-1} ({\tilde {\bv}} - {\tilde {\bv} }^{ref}) \right] {{\tilde J}}{d {\tilde \Omega}} \ = \nonumber \hspace{40mm} & & \\
{\rho} \int_{ \tilde \Omega } {\tilde {\bw} } \cdot {\tilde
{\bb}} {\tilde J} {d {\tilde \Omega}} \ + \ \int_{
\tilde \Omega } {\tilde p} {{{\tilde \nabla} \tilde {\bw} } : {\tilde
{\bF} }^{-t}} {\tilde J} {d {\tilde \Omega}} \ - \ 2
{\mu} \int_{ \tilde \Omega } {\tilde {\bD} }(\tilde {\bw} ) :
{\tilde {\bf D}}(\tilde {\bv}) {\tilde J} {d {\tilde
\Omega}} \nonumber & & \\
+  {\rho} \int_{ \tilde \Gamma } {\tilde {\bw}} {\tilde {\bP}} {\tilde
{\bN}} d {\tilde \Gamma} \hspace{60mm} & & 
\end{eqnarray}
\begin{eqnarray} \label{34}
\int_{\tilde \Omega} {\tilde q} { {{\tilde \nabla} \tilde {\bv} } :
{\tilde {\bF} }^{-t} } {d {\tilde \Omega}} &=& 0
\end{eqnarray}
is obtained, where $\tilde q$ and ${\tilde {\bw}}$ are respectively the
arbitrary pressure and velocity of the variational formulation. \renewcommand{\thefootnote}{\arabic{footnote}}

\subsection{Reconciling the Different Schools of Thought} \label{1001}

The equations (\ref{29}) and (\ref{30}) are the completely general
reference description of an incompressible, Newtonian fluid. They
reduce to the so--called A.L.E. equations of {\sc Hughes, Liu} and {\sc
Zimmerman} \cite{h:1} for an instant in which spatial and referential
configurations coincide. These simplified equations should, however,
not be implemented where the implementation requires evaluation about
more than one point within each time step (see Section \ref{44} for a
further, in--depth explanation). Under such circumstances the equations
of {\sc Hughes et al.} are an arbitrary Lagrangian Eulerian (A.L.E.)
description in the very true sense (this is not surprising considering
the equations have their origins in the arbitrarily, either Lagrangian
or Eulerian programmes of {\sc Hirt}, {\sc Amsden} and {\sc Cook}
\cite{h:5}). This fact is further borne out in observing that key
energetic properties, consistant with the principle of energy
conservation, are not automatically inherited by the equations of {\sc
Hughes et. al.} in the context of more general references.

The momentum equations of {\sc Soulaimani, Fortin, Dhatt} and {\sc
Ouellet} \cite{f:1} are flawed as a result of the mistaken belief that
${\tilde {\bsigma}} {\tilde {\bF}}^{-1} {\tilde J}$ is the
Piola--Kirchoff stress tensor of the first kind (pg. 268 of {\sc
Soulaimani et al.}). Yet another problem is illustrated by rewriting
the conventional incompressibility condition using the chain
rule. The new incompressibility condition which arises is most
certainly
\[
\frac{\partial \tilde{v}_i}{\partial \tilde{x}_j}\frac{\partial
\tilde{x}_j}{\partial x_i} = 0 \hspace{10mm} \mbox{and not}
\hspace{10mm} \frac{\partial \tilde{v}_i}{\partial
\tilde{x}_j}\frac{\partial \tilde{x}_i}{\partial x_j} = 0.
\]
Further errors arising (eg. $\hat J$ omitted in the first term on the
right hand side of the momentum equation, equation (10) on pg. 268 of
{\sc Soulaimani et al.}) make the use of these equations problematic.

\section{The Energetic Implications of a Deforming Reference}
\label{69}

The effect of quantities parameterising reference deformation on key
energetic properties -- nonlinear, exponential--type dissipation in the
absence of forcing and long--term stability under conditions of time
dependent loading -- is investigated in this section. These properties,
\[
K({\bv}) \le K({\bv} \mid_{t_0}) \ e^{-2 {\nu} C t} \hspace{10mm}
\mbox{and} \hspace{10mm} \lim_{t \rightarrow \infty} \sup K({\bv}) \le
\frac{M^2}{2 \nu^2 C^2}
\]
respectively (where $K = \frac{1}{2} {\rho} \left|\left| {\bv}
\right|\right|_{L^2(\Omega)}^2$ is the total kinetic energy), are
intrinsic to real flows and the conventional, Eulerian Navier--Stokes
equations (see {\sc Temam} \cite{Temam:1}, \cite{Temam:2}, {\sc
Constantin} and {\sc Foias} \cite{Constantin:1} and {\sc Simo} and {\sc
Armero} \cite{s:1} in this regard). The effect of ${\tilde
{\bv}}^{ref}$ on the afore mentioned aspects of conservation of the
quantity
\[
\frac{1}{2} {\rho} \left|\left| {\tilde {\bv}} {\tilde J}^{\frac{1}{2}}  \right|\right|_{L^2({\tilde \Omega})}^2
\]
is essentially what is being investigated, with a view to establishing
a set of conditions under which the discrete approximation can
reasonably be expected to inherit these self--same energetic
properties.

One might anticipate key energetic properties to be manifest only in
instances involving a fixed contributing mass of material, whether its
boundaries be dynamic, or not. An analysis of this nature only makes
sense in the context of a constant volume of fluid which, for
simplicity, will have material limits.

Inequalities of the Poincar\'e-Friedrichs type are a key feature of any
stability analysis of this nature. Gradient containing $L^2$ terms need
to be re--expressed in terms of energy. In the case of a ``no slip''
(${\bv} = 0$) condition on the entire boundary the situation is
straightforward, in that it is possible to use the standard
Poincare-Friedrichs inequality: there exists a constant $C_1 > 0$ such
that
\[
\|\bv\|_{L^2} \leq C_1 \|\nabla \bv\|_{L^2}\ \ \mbox{for all}\ \bv \in
[H^1_0(\Omega)]^n.
\]  
The use of the classical Poincar\'{e}--Friedrichs inequality is
otherwise identified as a major limitation, even in the conventional
Navier--Stokes related analyses. The Poincar\'{e}--Friedrichs
inequality is only applicable in very limited instances where the
value for the entire boundary is stipulated to be identically zero.
For boundary conditions of a more general nature, such as those
encountered in this study, in which parts of the boundary may be
either a free surface or subject to traction conditions, a more
suitable inequality is required (notice that subtracting a boundary
velocity and analising the resulting equation is not feasible as the
equations are nonlinear).  The Poincar\'{e}--Friedrichs inequality
does, furthermore, not hold on subdomains of the domain in question
and the constant is not optimal.

Further investigation ({\sc communication} \cite{comreddy:1}) reveals a
similar result, the so-called Poincar\'{e}--Morrey inequality, holds
providing the function attains a value of zero somewhere on the
boundary. The Poincar\'e-Morrey inequality states that a constant $C_2
> 0$ exists such that
\[
\|\bv\|_{L^2} \leq C_2 \|\nabla \bv\|_{L^2}\ \ \mbox{for all}\ \bv \in
[H^1_0(\Omega)]^n.
\]  
The proof of the Poincar\'e-Morrey inequality is, however, similar to
that of one of Korn's inequalities (see, for example, {\sc Kikuchi}
and {\sc Oden} \cite{kikuchi:1}). In particluar, it is
non--constructive, by contradiction and the constant cannot therefore
be determined as part of the proof. Viewed in this light the
forthcoming inequality amounts to a specification of the hypothetical
constant in the Poincar\'{e}--Morrey inequality for domains of a
particular geometry. The particular types of geometry considered are
those that arise in problems involving the motion of rigid bodies
such as pebbles on the sea bed; thus a free surface is present, and
the domain may be multiply connected.
   
\begin{inequality}[A New ``Poincar\'{e}'' Inequality] \label{94}
Suppose ${{\bv}}$ is continuous and differentiable to first order and
that ${{\bv}}$ attains a maximum absolute value, $c$, on an included,
finite neighbourhood of minimum radius $R_{\mbox{\it \scriptsize min}}$
about a point ${{\bx}}^{\mbox{\scriptsize origin}}$ (as depicted in
Figure \ref{194}).
\begin{figure}[H]
\begin{center} \leavevmode
\mbox{\epsfbox{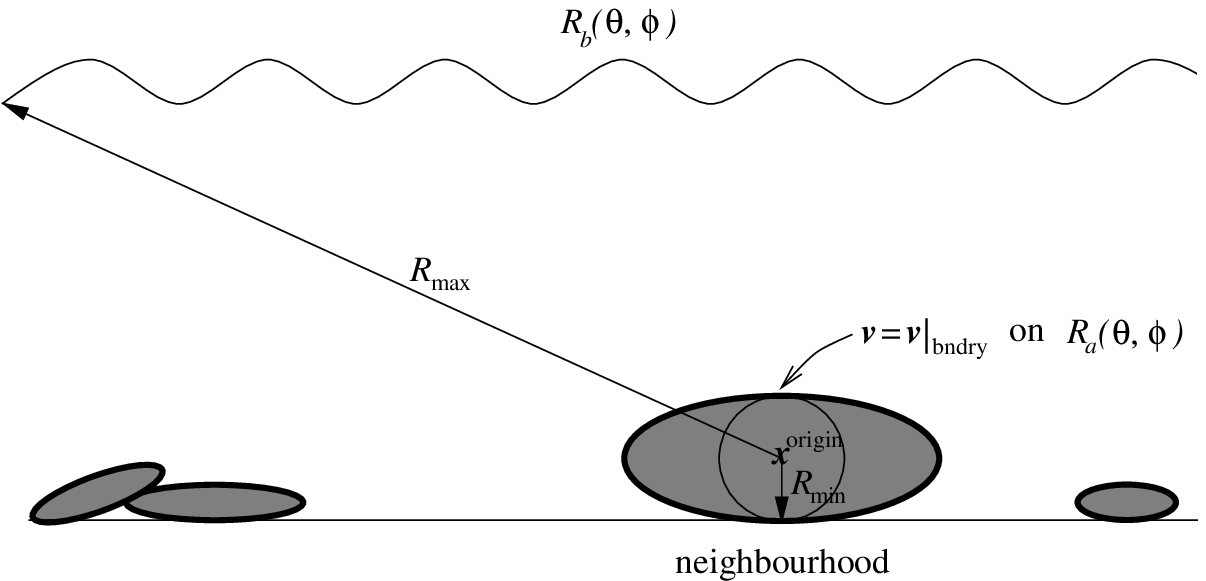}}
\end{center}
\caption{A Finite Neighbourhood of Minimum Radius $R_{\mbox{\it
\scriptsize min}}$ About a Point ${{\bx}}^{\mbox{\scriptsize
origin}}$.} \label{194}
\end{figure} 
If ${\Omega}$ is a bounded, star--shaped (about a point
${{\bx}}^{\mbox{\scriptsize origin}}$)\footnotemark[1] domain in $R^3$,
then \footnotetext[1]{by which is meant that every point in the domain
can be reached by a straight line from ${{\bx}}^{\mbox{\scriptsize
origin}}$ that does not pass outside of $\Omega$}
\[
\left|\left| {\bv} \right|\right|_{L^2({ \Omega})} \le { \left[
\frac{(R_{\mbox{\it \scriptsize max}} - R_{\mbox{\it \scriptsize min}})
(R_{\mbox{\it \scriptsize max}}^3 - R_{\mbox{\it \scriptsize min}}^3)}
{3 R_{\mbox{\it \scriptsize max}} R_{\mbox{\it \scriptsize min}}}
\right]^{\frac{1}{2}}} \left|\left| { \nabla} { {\bv}}
\right|\right|_{L^2({ \Omega})} + \left|\left| c
\right|\right|_{L^2({\Omega})}
\]
where $R_{\mbox{\it \scriptsize max}}$ is the distance to the farthest
point in ${ \Omega}$ from ${{\bx}}^{\mbox{\scriptsize origin}}$.
\end{inequality}

{\sc Proof:} Consider the change to spherical coordinates
\[
{\breve v}_i(r, \theta, \phi) = { v}_i(r \mathop{\rm sin}\theta
\mathop{\rm cos}\phi - { x}_1^{\mbox{\scriptsize origin}}, r
\mathop{\rm sin}\theta \mathop{\rm sin}\phi - {
x}_2^{\mbox{\scriptsize origin}}, r \mathop{\rm cos}\theta -
{ x}_3^{\mbox{\scriptsize origin}})
\]
centred on ${{\bx}}^{\mbox{\scriptsize origin}}$. Suppose the radial
limits of the domain and neighbourhood are denoted $R_b (\theta, \phi)$
and $R_a (\theta, \phi)$ respectively. By the fundamental theorem of
integral calculus
\begin{eqnarray*}
\left( {\breve v}_i(r, \theta, \phi) - {\breve v}_i\mid_{R_a (\theta,
\phi)} \right)^2 &=& \left( \int_{R_a (\theta, \phi)}^r \frac{\partial
{\breve v}_i}{\partial r} (\xi, \theta, \phi) d \xi \right)^2 \\ 
&& \\
&=& \left( \int_{R_a (\theta, \phi)}^{r} \frac{1}{\xi} \xi
\frac{\partial {\breve v}_i}{\partial r} (\xi, \theta, \phi) d \xi
\right)^2 \\
&& \\
&\le& \int_{R_a (\theta, \phi)}^r \frac{1}{\xi^2} d \xi \int_{R_a
(\theta, \phi)}^r \left( \frac{\partial {\breve v}_i}{\partial r}(\xi,
\theta, \phi) \right)^2 \xi^2 d \xi \\
&& \hspace{10mm} \mbox{\it (by Schwarz inequality)} \\
&\le& \int_{R_{\mbox{\it \scriptsize min}}}^{R_{\mbox{\it \scriptsize
max}}} \frac{1}{\xi^2} d \xi \int_{R_a (\theta, \phi)}^{R_b (\theta,
\phi)} \left( \frac{\partial {\breve v}_i}{\partial {r}}(\xi,
\theta, \phi) \right)^2 \xi^2 d \xi \hspace{10mm} \mbox{\it (for } r
\in {\breve \Omega} \mbox{\it)} \\
&& \\
&=& \frac{(R_{\mbox{\it \scriptsize max}} - R_{\mbox{\it \scriptsize
min}})}  {R_{\mbox{\it \scriptsize max}} R_{\mbox{\it \scriptsize
min}}} \int_{R_a (\theta, \phi)}^{R_b (\theta, \phi)} \left(
\frac{\partial {\breve v}_i}{\partial r}(\xi, \theta, \phi)
\right)^2 \xi^2 d \xi \\
&& \\ 
&=& \frac{(R_{\mbox{\it \scriptsize max}} - R_{\mbox{\it \scriptsize
min}})}  {R_{\mbox{\it \scriptsize max}} R_{\mbox{\it \scriptsize
min}}} {\breve V}_i (\theta, \phi)
\end{eqnarray*}
\[
\mbox{where} \hspace{10mm} {\breve V}_i (\theta, \phi) = \int_{R_a
(\theta, \phi)}^{R_b (\theta, \phi)} \left( \frac{\partial {\breve
v}_i}{\partial r}(\xi, \theta, \phi) \right)^2 \xi^2 d \xi.
\]
Integrating this result over that part of ${\breve \Omega}$ outside the
neighbourhood (angular extent being $\Theta_a(\phi) \le \theta \le
\Theta_b(\phi)$ and $\Phi_a \le \phi \le \Phi_b$)
\begin{eqnarray*}
&& \hspace{-7mm} \int_{\Phi_a}^{\Phi_b}
\int_{\Theta_a(\phi)}^{\Theta_b(\phi)} \int_{R_a (\theta, \phi)}^{R_b
(\theta, \phi)} \left( {\breve v}_i(r, \theta, \phi) - {\breve
v}_i\mid_{R_a (\theta, \phi)} \right)^2 r^2 \mathop{\rm sin}\theta d r
d \theta d \phi \\
&& \hspace{13mm} \le \frac{(R_{\mbox{\it \scriptsize max}} -
R_{\mbox{\it \scriptsize min}})}  {R_{\mbox{\it \scriptsize max}}
R_{\mbox{\it \scriptsize min}}} \int_{\Phi_a}^{\Phi_b}
\int_{\Theta_a(\phi)}^{\Theta_b(\phi)} \int_{R_a (\theta, \phi)}^{R_b
(\theta, \phi)} {\breve V}_i (\theta, \phi) r^2 \mathop{\rm sin}\theta
d r d \theta d \phi \\
&& \\
&& \hspace{13mm} \le \frac{(R_{\mbox{\it \scriptsize max}} -
R_{\mbox{\it \scriptsize min}})}  {R_{\mbox{\it \scriptsize max}}
R_{\mbox{\it \scriptsize min}}} \int_{\Phi_a}^{\Phi_b}
\int_{\Theta_a(\phi)}^{\Theta_b(\phi)} {\breve V}_i (\theta, \phi)
\left( \int_{R_{\mbox{\it \scriptsize min}}}^{R_{\mbox{\it \scriptsize
max}}} r^2 d r \right) \mathop{\rm sin}\theta d \theta d \phi \frac{}{}
\\
&& \\
&& \hspace{13mm} \le \frac{(R_{\mbox{\it \scriptsize max}} -
R_{\mbox{\it \scriptsize min}}) (R_{\mbox{\it \scriptsize max}}^3 -
R_{\mbox{\it \scriptsize min}}^3)} {3 R_{\mbox{\it \scriptsize max}}
R_{\mbox{\it \scriptsize min}}} \int_{\Phi_a}^{\Phi_b}
\int_{\Theta_a(\phi)}^{\Theta_b(\phi)} \int_{R_a (\theta, \phi)}^{R_b
(\theta, \phi)}  \left( \frac{\partial {\breve v}_i}{\partial r}
\right)^2 {r}^2 \mathop{\rm sin}\theta d r d \theta d \phi \\
&& \\
&& \hspace{13mm} \le \frac{(R_{\mbox{\it \scriptsize max}} -
R_{\mbox{\it \scriptsize min}}) (R_{\mbox{\it \scriptsize max}}^3 -
R_{\mbox{\it \scriptsize min}}^3)} {3 R_{\mbox{\it \scriptsize max}}
R_{\mbox{\it \scriptsize min}}} \int_{\Phi_a}^{\Phi_b}
\int_{\Theta_a(\phi)}^{\Theta_b(\phi)} \int_{R_a (\theta, \phi)}^{R_b
(\theta, \phi)} \left[ \left( \frac{\partial {\breve v}_i}{\partial r}
\right)^2 + \frac{1}{r^2} \left( \frac{\partial {\breve v}_i}{\partial
\theta} \right)^2 \right. \\ && \hspace{83mm} \left. + \frac{1}{r^2
\sin^2 \theta} \left( \frac{\partial {\breve v}_i}{\partial
\phi} \right)^2 \right] r^2 \mathop{\rm sin}\theta d r d \theta d \phi
\\
&& \\
&& \hspace{13mm} = \frac{(R_{\mbox{\it \scriptsize max}} - R_{\mbox{\it
\scriptsize min}}) (R_{\mbox{\it \scriptsize max}}^3 - R_{\mbox{\it
\scriptsize min}}^3)} {3 R_{\mbox{\it \scriptsize max}} R_{\mbox{\it
\scriptsize min}}} \int_{\Phi_a}^{\Phi_b}
\int_{\Theta_a(\phi)}^{\Theta_b(\phi)} \int_{R_a (\theta, \phi)}^{R_b
(\theta, \phi)} \left( \nabla {\breve v}_i \right) \cdot \left( \nabla
{\breve v}_i \right) r^2 \mathop{\rm sin}\theta dr d \theta d \phi.
\end{eqnarray*}

Changing back to the original rectangular coordinates and defining
${{\bv}}\mid_{\mbox{\it \scriptsize bndry}}$ to be a radially constant
function throughout $\Omega$ which takes the values of ${\breve {\bv}}
\mid_{R_a (\theta, \phi)}$ for $r = R_a(\theta, \phi)$,
\begin{eqnarray*}
\int_{\Omega_*} \left( { v}_i({ {\bx}}) - { v}_i\mid_{\mbox{\it
\scriptsize bndry}} \right)^2 d { \Omega} &\le& \frac{(R_{\mbox{\it
\scriptsize max}} - R_{\mbox{\it \scriptsize min}}) (R_{\mbox{\it
\scriptsize max}}^3 - R_{\mbox{\it \scriptsize min}}^3)} {3
R_{\mbox{\it \scriptsize max}} R_{\mbox{\it \scriptsize min}}}
\int_{\Omega_*} \left( \nabla { v}_i({ {\bx}}) \right) \cdot \left(
\nabla { v}_i({ {\bx}}) \right) d {\Omega}
\end{eqnarray*}
where $\Omega_*$ is $\Omega$ excluding the neighbourhood. Summing over
$i$,
\begin{eqnarray*}
\int_{\Omega_*} \left( { {\bv}} - {\bv}\mid_{\mbox{\it \scriptsize
bndry}} \right) \cdot \left( { {\bv}} - { {\bv}}\mid_{\mbox{\it
\scriptsize bndry}} \right) d { \Omega} &\le& \frac{(R_{\mbox{\it
\scriptsize max}} - R_{\mbox{\it \scriptsize min}}) (R_{\mbox{\it
\scriptsize max}}^3 - R_{\mbox{\it \scriptsize min}}^3)} {3
R_{\mbox{\it \scriptsize max}} R_{\mbox{\it \scriptsize min}}}
\int_{\Omega_*} \left( \nabla { {\bv}} \right) : \left( \nabla { {\bv}}
\right) d {\Omega}.
\end{eqnarray*}
Making use of either the Cauchy--Schwarz or triangle inequality, 
\begin{eqnarray*}
\left( \left|\left| {\bv} \right|\right|_{L^2({\Omega_*})} - \left|\left|
{\bv}\mid_{\mbox{\it \scriptsize bndry}} \right|\right|_{L^2({\Omega_*})}
\right)^2 &\le& \frac{(R_{\mbox{\it \scriptsize max}} - R_{\mbox{\it
\scriptsize min}}) (R_{\mbox{\it \scriptsize max}}^3 - R_{\mbox{\it
\scriptsize min}}^3)} {3 R_{\mbox{\it \scriptsize max}} R_{\mbox{\it
\scriptsize min}}} \left|\left| \nabla {\bv} \right|\right|_{L^2({\Omega_*})}^2,
\end{eqnarray*}
and remembering that $\sup \mid {\breve {\bv}} \mid_{R_a (\theta,
\phi)} \mid \le c$,
\begin{eqnarray*}
\left|\left| {\bv} \right|\right|_{L^2({\Omega_*})} &\le& \left[
\frac{(R_{\mbox{\it \scriptsize max}} - R_{\mbox{\it \scriptsize min}})
(R_{\mbox{\it \scriptsize max}}^3 - R_{\mbox{\it \scriptsize min}}^3)}
{3 R_{\mbox{\it \scriptsize max}} R_{\mbox{\it \scriptsize min}}}
\right]^{\frac{1}{2}} \left|\left| { \nabla}{ {\bv}}
\right|\right|_{L^2({\Omega_*})} + \left|\left| c \right|\right|_{L^2({\Omega_*})}.
\end{eqnarray*}
Consider the terms $\left|\left| {\bv} \right|\right|_{L^2}$ and
$\left|\left| c \right|\right|_{L^2}$. Comparing these terms under
circumstances of $\sup \mid {\bv} \mid \le c$ leads to the conclusion
that the inequality holds over the neighbourhood and that the
inequality is therefore unaffected when the domain of integration is
extended to include the neighbourhood. Of course, the radial extension
of ${\bv}\mid_{\mbox{\it \scriptsize bndry}}$ can be used in place of
$c$ in instances where inclusion of the neighbourhood is not required.

This inequality is similar to the Poincar\'{e}--Friedrichs inequality
when $c=0$, but is extended to a geometrical subclass of domains which
have free and partly non-zero boundaries. It has a further advantage in
that the constant is an order of magnitude more optimal when used under
the ``no slip'' Poincar\'{e}--Friedrichs condition (under such
conditions the domain can always be deconstructed into a number of
subdomains in which $R_{\mbox{\it \scriptsize min}} =
\frac{1}{3}R_{\mbox{\it \scriptsize max}}$). The
Poincar\'{e}--Friedrichs inequality is a special case of the above
inequality. The necessary lemma (below) follows naturally from the
above inequality.

\begin{lemma}[Deviatoric Stress Term Energy] \label{58}
The kinetic energy satisfies the bound $ \ \displaystyle \frac{C}{\rho}
{\tilde K}({\tilde {\bv}}) \ \le \ \left|\left| {\tilde {\bD}}({\tilde
{\bv}}) {\tilde J}^{\frac{1}{2}} \right|\right|_{L^2({\tilde
\Omega})}^2$, where $C$ is related to the constant in Inequality
\ref{94}, $C > 0$.
\end{lemma}
{\sc Proof:} If, in particular, ${ {\bv}}\mid_{\mbox{\it \scriptsize
bndry}} = 0$ in Inequality \ref{94},
\begin{eqnarray*}
\left|\left| { {\bv}} \right|\right|_{L^2({ \Omega})} &\le&
\frac{\left|\left| { \nabla} { {\bv}} \right|\right|_{L^2({
\Omega})}}{\sqrt {C}} \\
C\frac{1}{2}\left|\left| { {\bv}} \right|\right|_{L^2({ \Omega})}^2
&\le& \left|\left| { {\bD}}({ {\bv}}) \right|\right|_{L^2({ \Omega})}^2
\end{eqnarray*}
(The relationship between ${\bD}$ and ${\nabla} {\bv}$ arises in the
context of the original equations involving $\mathop{\rm
div}{\bsigma}$. It is because \begin{eqnarray*}
\begin{array}{ccll} D_{ij,j} &=& \frac{1}{2} \left( v_{i,jj} + v_{j,ij}
\right) & \\
&=& \frac{1}{2} \left( v_{i,jj} + v_{j,ji} \right)
\hspace{10mm} & \mbox{\em (changing the order of differentiation)} \\
&=& \frac{1}{2} v_{i,jj} & \mbox{\em (${\mathop {\rm div}}{\bv} = 0$ by
incompressibility}), \end{array} \end{eqnarray*}
assuming, of course, that ${\bv}$ is continuous and differentiable to
first order.) Rewriting in terms of ${\tilde \Omega}$
\begin{eqnarray*}
\frac{C}{\rho} {\tilde K}({\tilde {\bv}}) &\le& \left|\left| {\tilde
{\bD}}({\tilde {\bv}}) {\tilde J}^{\frac{1}{2}}
\right|\right|_{L^2({\tilde \Omega})}^2.
\end{eqnarray*}

The following lemma is vital to the deforming reference analysis in
particular. It will form the basis to the next lemma and another (on
page \pageref{354}) concerned with the time discrete analysis.

\begin{lemma} [Basic to Lemmas \ref{355} and \ref{354}] \label{54} 
The relation 
\[
\left<{\tilde{\bu}}, ({\tilde \nabla} {\tilde {\bv}})
{\tilde {\bF}}^{-1} {\tilde {\bw}} {\tilde J}  \right>_{L^2(\tilde
\Omega)} = - \left< {\tilde {\bv}}, ({\tilde \nabla}{\tilde{\bu}})
{\tilde {\bF}}^{-1} {\tilde {\bw}} {\tilde J} \right>_{L^2(\tilde
\Omega)} - \left< {\tilde{\bu}} \left( {\tilde \nabla}{\tilde
{\bw}}:{\tilde {\bF}}^{-t} \right), {\tilde {\bv}} {\tilde J}
\right>_{L^2(\tilde \Omega)}
\] 
is valid for
\[
{\tilde {\bw}} \in W = \left\{ {\tilde{\bw}} \ : \ {\tilde {\bw}} =
{\bf 0} \mbox{ or } {\tilde {\bF}}^{-t}{\tilde {\bN}} \cdot {\tilde
{\bw}} = 0 \mbox{ on } {\tilde \Gamma} \right\}.
\]
\end{lemma}
{\sc Proof:} Consider ${\tilde{\bu}} \cdot ({\tilde \nabla}
{\tilde{\bv}}) {\tilde {\bF}}^{-1} {\tilde {\bw}} {\tilde J}$:
\begin{eqnarray*}
{\tilde u}_i {\tilde v}_{i,j} {\tilde F}^{-1}_{jk}{\tilde w}_k {\tilde
J} &=& - {\tilde u}_{i,j} {\tilde v}_i {\tilde F}^{-1}_{jk} {\tilde
w}_k {\tilde J} - {\tilde u}_i {\tilde v}_i ({\tilde F}^{-1}_{jk}
{\tilde w}_k {\tilde J})_{,j} + ({\tilde u}_i {\tilde v}_i {\tilde
F}^{-1}_{jk} {\tilde w}_k {\tilde J})_{,j}
\end{eqnarray*}
by the product rule. In the terms arising from $({\tilde F}^{-1}_{jk}
{\tilde w}_k {\tilde J})_{,j}$, both ${\tilde F}^{-1}_{jk,j}$ and ${\tilde J}_{,j} {\tilde F}^{-1}_{jk}$ vanish under the condtions specified (in section \ref{1001}) for equations of {\sc Hughes, Liu} and {\sc
Zimmerman} \cite{h:1} to be a completely general reference description. Thus
\begin{eqnarray*}
{\tilde u}_i {\tilde v}_{i,j} {\tilde F}^{-1}_{jk} {\tilde w}_k {\tilde
J} &=& - {\tilde u}_{i,j} {\tilde v}_i {\tilde F}^{-1} _{jk} {\tilde
w}_k {\tilde J} - {\tilde u}_i {\tilde v}_i {\tilde F}^{-1}_{jk}
{\tilde w}_{k,j} {\tilde J} + ({\tilde u}_i {\tilde v}_i {\tilde
F}^{-1} _{jk} {\tilde w}_k {\tilde J})_{,j}.
\end{eqnarray*} 
Integrating over the domain ${\tilde {\Omega}}$ and applying the
divergence theorem,
\begin{eqnarray} \label{309}
\left<{\tilde{\bu}}, ({\tilde \nabla} {\tilde {\bv}}) {\tilde
{\bF}}^{-1} {\tilde {\bw}} {\tilde J}  \right>_{L^2(\tilde
\Omega)} &=& - \left< {\tilde {\bv}}, ({\tilde \nabla}{\tilde{\bu}})
{\tilde {\bF}}^{-1} {\tilde {\bw}} {\tilde J} 
\right>_{L^2(\tilde \Omega)} \nonumber \\
&& - \left< {\tilde{\bu}} \left( {\tilde \nabla}{\tilde {\bw}}:{\tilde
{\bF}}^{-t} \right), {\tilde {\bv}} {\tilde J} \right>_{L^2(\tilde
\Omega)} \nonumber \\
&& + \left< {\tilde{\bu}}, {\tilde {\bv}} \left( {\tilde {\bF}}^{-t}
{\tilde {\bN}} \cdot {\tilde {\bw}} \right) {\tilde J} 
\right>_{L^2(\tilde \Gamma)} 
\end{eqnarray}
The condition of this lemma dictates the manner in which the reference
must deform to ensure that the equation will inherit the desired
energetic properties. This lemma is crucial to the deforming reference
analysis. The lemma immediately below will facilitate the elimination
of the convective energy rate in the forthcoming analysis.

\begin{lemma} [Convective Energy Rate] \label{355} The relation
\begin{eqnarray*}
- {\rho} \left< {\tilde {\bv}}, ({\tilde \nabla} {\tilde {\bv}})
{\tilde {\bF}}^{-1} \left( {\tilde {\bv}} - {\tilde
{\bv}}^{\mbox{\scriptsize {\em ref}}} \right) {\tilde J}
\right>_{L^2({\tilde \Omega})} &=& - \frac{1}{2} \rho \left< {\tilde
{\bv}}, {\tilde {\bv}} \frac{\partial {\tilde J}}{\partial t}
\right>_{L^2({\tilde \Omega})}
\end{eqnarray*}
is valid in instances where a purely Lagrangian description is used to
track free boundaries and/or boundaries are of a fixed impermeable
type.
\end{lemma}
{\sc Proof:} In instances where a purely Lagrangian description is used
to track free boundaries, ${\tilde {\bv}} - {\tilde
{\bv}}^{\mbox{\scriptsize {\em ref}}}$ vanishes, as does ${\tilde
{\bF}}^{-t}{\tilde {\bN}}\cdot{\tilde {\bv}}$ at fixed impermeable
boundaries. The condition at the boundary for Lemma \ref{54} is
therefore satisfied.  Thus the term
\begin{eqnarray*} \label{310}
- {\rho} \left< {\tilde {\bv}}, ({\tilde \nabla} {\tilde {\bv}})
{\tilde {\bF}}^{-1} \left( {\tilde {\bv}} - {\tilde
{\bv}}^{\mbox{\scriptsize {\em ref}}} \right) {\tilde J}
\right>_{L^2({\tilde \Omega})} &=& {\rho} \left< {\tilde {\bv}},
({\tilde \nabla}{\tilde{\bv}}) {\tilde {\bF}}^{-1} \left( {\tilde
{\bv}} - {\tilde {\bv}}^{\mbox{\scriptsize {\em ref}}} \right) {\tilde
J} \right>_{L^2(\tilde \Omega)} \nonumber \\
&& + {\rho} \left< {\tilde{\bv}} \left( {\tilde \nabla}\left( {\tilde
{\bv}} - {\tilde {\bv}}^{\mbox{\scriptsize {\em ref}}} \right):{\tilde
{\bF}}^{-t} \right), {\tilde {\bv}} {\tilde J} \right>_{L^2(\tilde
\Omega)} \nonumber \\
&& \nonumber \\
&=& {\rho} \left< {\tilde {\bv}}, ({\tilde \nabla}{\tilde{\bv}})
{\tilde {\bF}}^{-1} \left( {\tilde {\bv}} - {\tilde
{\bv}}^{\mbox{\scriptsize {\em ref}}} \right) {\tilde J}
\right>_{L^2(\tilde \Omega)} \nonumber \\
&& - {\rho} \left< {\tilde{\bv}} \left( {\tilde \nabla}{\tilde
{\bv}}^{\mbox{\scriptsize {\em ref}}}:{\tilde {\bF}}^{-t} \right),
{\tilde {\bv}} {\tilde J} \right>_{L^2(\tilde \Omega)} \hspace{5mm}
\mbox{\em (by incomp--} \nonumber \\
&& \hspace{62mm} \mbox{\em ressibility)} \nonumber \\
&=& - \frac{1}{2} {\rho} \left< {\tilde{\bv}} \left( {\tilde
\nabla}{\tilde {\bv}}^{\mbox{\scriptsize {\em ref}}}:{\tilde
{\bF}}^{-t} \right), {\tilde {\bv}} {\tilde J} \right>_{L^2(\tilde
\Omega)} \nonumber \\
&& \nonumber \\
&=& - \frac{1}{2} \rho \left< {\tilde {\bv}}, {\tilde {\bv}}
\frac{\partial {\tilde J}}{\partial t} \right>_{L^2({\tilde \Omega})}
\nonumber
\end{eqnarray*}
since $\displaystyle \frac{\partial {\tilde J}}{\partial t} = {\tilde
J} div {\bv}^{\scriptsize ref}$ (which is ${\tilde J}{\tilde
\nabla}{\tilde {\bv}}^{\mbox{\scriptsize {\em ref}}}:{\tilde
{\bF}}^{-t}$) in the same vein as ${\dot {\cal J}}_0 = {\cal J}_0
\mathop{\rm div}{\bv}$ (the kinematic result on used earlier).

This lemma concludes the preliminaries required for the deforming
reference energy analysis.

\subsection{Exponential Dissipation in the Absence of Forcing}

The issue of whether nonlinear, exponential--type dissipation in the
absence of forcing is a property intrinsic to the deforming reference
description is resolved as follows.

\begin{theorem}[Exponential Dissipation in the Absence of Forcing]
\label{60}
A sufficient condition for the completely general reference
description to inherit nonlinear, exponential type energy dissipation
\[
{\tilde K}({\tilde {\bv}}) \le {\tilde K}({\tilde {\bv}} \mid_{t_0})
\ e^{-2 {\nu} C t} 
\]
(where ${\tilde K}({\tilde {\bv}}) \equiv \displaystyle \frac{1}{2}
{\rho} \left|\left| {\tilde {\bv}} {\tilde J}^{\frac{1}{2}}
\right|\right|_{L^2({\tilde \Omega})}^2$)
in the absence of forcing (an intrinsic feature of real flows and the
conventional, Eulerian Navier--Stokes equations) is that the reference
moves in a purely Lagrangian fashion at free boundaries.
\end{theorem}
{\sc Proof:} The first step towards formulating an expression involving
the kinetic energy is to substitute ${\tilde {\bv}}$ for ${\tilde
{\bw}}$ in the variational momentum equation (\ref{33}) on page
\pageref{33}. Then
\begin{eqnarray} \label{74}
{\rho} \left< {\tilde {\bv}}, \frac{\partial {\tilde {\bv}}}{\partial
t} {\tilde J} \right>_{L^2(\tilde \Omega)} &=& \left< {\tilde p}
{\tilde \nabla} {\tilde {\bv}} , {\tilde {\bF}}^{-t} {\tilde J}
\right>_{L^2({\tilde \Omega})} - 2 {\mu} \left< {\tilde {\bD}}({\tilde
{\bv}}), {\tilde {\bD}}({\tilde {\bv}}) {\tilde J} \right>_{L^2(\tilde
\Omega)} \nonumber \\
&& - {\rho} \left< {\tilde {\bv}}, ({\tilde \nabla} {\tilde {\bv}})
{\tilde {\bF}}^{-1} \left( {\tilde {\bv}} - {\tilde
{\bv}}^{\mbox{\scriptsize {\em ref}}} \right) {\tilde J}
\right>_{L^2({\tilde \Omega})} \nonumber \\
&& + {\rho} \left< {\tilde {\bv}}, {\tilde {\bb}} {\tilde J}
\right>_{L^2({\tilde \Omega})} + \left< {\tilde{\bv}}, {\tilde{\bP}}
{\tilde {\bN} } \right>_{L^2({\tilde \Gamma})}.
\end{eqnarray}
The term containing the pressure, that is 
\[
\left< {\tilde p} {\tilde \nabla} {\tilde {\bv}} :  {\tilde {\bF}}^{-t}
{\tilde J} \right>_{L^2({\tilde \Omega})},
\]
vanishes as a result of incompressibility (equation (\ref{30})). The order of integration and differentiation are
interchangeable (limits are time--independent in the reference which
tracks the free boundary perfectly -- a description which becomes fully
Lagrangian at boundaries was stipulated). Equation (\ref{74}) can be
rewritten
\begin{eqnarray*} 
\frac{1}{2}{\rho} \left( \frac{d}{dt} \left|\left| {\tilde {\bv}}
{\tilde J}^{\frac{1}{2}}  \right|\right|_{L^2({\tilde \Omega})}^2 -
\left< {\tilde {\bv}}, {\tilde {\bv}} \frac{\partial {\tilde
J}}{\partial t} \right>_{L^2({\tilde \Omega})} \right) &=& -2 {\mu}
\left|\left| {\tilde {\bD}} {\tilde J}^{\frac{1}{2}}
\right|\right|_{L^2({\tilde \Omega})}^2 \\
&& - {\rho} \left< {\tilde {\bv}}, ({\tilde \nabla} {\tilde {\bv}})
{\tilde {\bF}}^{-1} \left( {\tilde {\bv}} - {\tilde
{\bv}}^{\mbox{\scriptsize {\em ref}}} \right) {\tilde J}
\right>_{L^2({\tilde \Omega})} \\
&& + {\rho} \left< {\tilde {\bv}}, {\tilde {\bb}} {\tilde J}
\right>_{L^2(\tilde \Omega)} + \left< {\tilde {\bv}}, {\tilde
{\bP}}{\tilde {\bN}} \right>_{L^2(\tilde \Gamma)}
\end{eqnarray*}
as a result. The conditions of Lemma \ref{355} are also satisfied for a
description which becomes fully Lagrangian at free boundaries and an
expression
\begin{eqnarray*}
\frac{d {\tilde K}({\tilde {\bv}})}{dt} &=& -2 {\mu} \left|\left|
{\tilde {\bD}} {\tilde J}^{\frac{1}{2}}  \right|\right|_{L^2({\tilde
\Omega})}^2 + {\rho} \left< {\tilde {\bv}}, {\tilde {\bb}} {\tilde J}
\right>_{L^2(\tilde \Omega)} + \left< {\tilde {\bv}}, {\tilde
{\bP}}{\tilde {\bN}} \right>_{L^2(\tilde \Gamma)}
\end{eqnarray*}
is therefore obtained, where ${\tilde K} = \displaystyle \frac{1}{2}
{\rho} \left|\left| {\tilde {\bv}} {\tilde J}^{\frac{1}{2}} 
\right|\right|_{L^2({\tilde \Omega})}^2$ is the total kinetic energy.
Using Lemma \ref{58}
\begin{eqnarray} \label{59}
\frac{d {\tilde K}({\tilde {\bv}})}{dt} &\le& -2 {\nu} C {\tilde
K}({\tilde {\bv}}) + {\rho} \left< {\tilde {\bv}}, {\tilde {\bb}}
{\tilde J} \right>_{L^2(\tilde \Omega)} + \left< {\tilde {\bv}},
{\tilde {\bP}}{\tilde {\bN}} \right>_{L^2(\tilde \Gamma)}.
\end{eqnarray}
Equation (\ref{59}) has  a solution of the form
\[
{\tilde K} \le {\tilde K}({\tilde {\bv}} \mid_{t_0}) \ e^{-2 {\nu} C t}
\]
in the absence of forcing ($\mbox{\it ``no forcing''} \Rightarrow
{\tilde {\bb}} = {\tilde {\bP}}{\tilde {\bN}} = {\bf 0}$), providing a
purely Lagrangian description is used at free boundaries.
 
A nonlinear, exponential--type energy dissipation in the absence of
forcing is therefore an intrinsic property of the completely general
reference description. This contractive flow property is also an
intrinsic property of the conventional Navier--Stokes equations.

\subsection{Long--Term Stability under Conditions of Time--Dependent
Loading}

The formulation of suitable load and free surface bounds is necessary
before the issue of long-term stability ($L^2$--stability) under
conditions of time--dependent loading can be resolved. The following
lemma facilitates the formulation of load and free surface bounds.

\begin{lemma}[Force, Free Surface Bounds] \label{49} The inequality
\begin{eqnarray*}
{\rho} \left< {\tilde {\bv}}, {\tilde {\bb}} {\tilde J}
\right>_{L^2({\tilde \Omega})} + \left< {\tilde {\bv}}, {\tilde {\bP}}
{\tilde {\bN}} \right>_{L^2(\tilde \Gamma)} &\le& \displaystyle
\frac{\nu C}{2} \left( {\rho} \left|\left| {\tilde {\bv}} {\tilde
J}^{\frac{1}{2}} \right|\right|_{L^2({\tilde \Omega})}^2 + \left|\left|
{\tilde {\bv}} \right|\right|_{L^2({\tilde \Gamma})}^2 \right) \\
&& + \frac{1}{2 \nu C} \left( {\rho} \left|\left| {\tilde {\bb}}
{\tilde J}^{\frac{1}{2}} \right|\right|_{L^2(\tilde \Omega)}^2 +
\left|\left| {\tilde {\bP}} {\tilde {\bN}} \right|\right|_{L^2(\tilde
\Gamma)}^2 \right)
\end{eqnarray*}
holds where $\nu C$ is a constant, $\nu C > 0$.
\end{lemma}
{\sc Proof:} In terms of the Cauchy--Schwarz inequality,
\begin{eqnarray*} 
\left< {\tilde {\bv}}, {\tilde {\bb}} {\tilde J} \right>_{L^2({\tilde
\Omega})} &\le& \left|\left| {\tilde {\bv}} {\tilde J}^{\frac{1}{2}}
\right|\right|_{L^2({\tilde \Omega})} \ \left|\left| {\tilde {\bb}}
{\tilde J}^{\frac{1}{2}} \right|\right|_{L^2(\tilde \Omega)} \\
&\le& \frac{\nu C}{2} \left|\left| {\tilde {\bv}} {\tilde
J}^{\frac{1}{2}} \right|\right|_{L^2(\tilde \Omega)}^2 + \frac{1}{2 \nu
C} \left|\left| {\tilde {\bb}} {\tilde J}^{\frac{1}{2}}
\right|\right|^2_{L^2(\tilde \Omega)} \hspace{5mm} \mbox{for }
\hspace{5mm} \nu C > 0
\end{eqnarray*}
by Young's inequality. Similarly,
\begin{eqnarray*}
\left< {\tilde {\bv}}, {\tilde {\bP}} {\tilde {\bN}}
\right>_{L^2(\tilde \Gamma)} &\le& \frac{\nu C}{2} \left|\left| {\tilde
{\bv}} \right|\right|_{L^2({\tilde \Gamma})}^2 + \frac{1}{2 \nu C}
\left|\left| {\tilde {\bP}} {\tilde {\bN}} \right|\right|^2_{L^2(\tilde
\Gamma)} \hspace{5mm} \mbox{for } \hspace{5mm} \nu C > 0.
\end{eqnarray*}

This done, the mathematical machinery necessary to the long--term
stability analysis is in place.

\begin{theorem}[Long--Term Stability] \label{61} 
A sufficient condition for the completely general reference
description to inherit the property of long--term stability
\[
\lim_{t \rightarrow \infty} \sup {\tilde K}({\tilde {\bv}}) \le
\frac{M^2}{2 \nu^2 C^2}
\] 
under conditions of time--dependent loading (an intrinsic feature of
real flows and the Navier--Stokes equations), where this
time--dependent loading and the speed of the free surface is bounded in
such a way that
\[
{\rho} \left|\left| {\tilde {\bb}} {\tilde J}^{\frac{1}{2}}  \right|\right|_{L^2(\tilde \Omega)}^2 + \left|\left|
{\tilde {\bP}}{\tilde {\bN}} \right|\right|_{L^2({\tilde \Gamma})}^2 +
\nu^2 C^2 \left|\left| {\tilde {\bv}} \right|\right|_{L^2({\tilde
\Gamma})}^2 \le M^2,
\]
is that the description becomes purely Lagrangian at free boundaries.
\end{theorem}
{\sc Proof:}
Using Lemma \ref{49} in equation (\ref{59}), then applying the above
bound,
\begin{eqnarray*}
\frac{d {\tilde K}({\tilde {\bv}})}{dt} + \nu C {\tilde K}({\tilde
{\bv}}) &\le& \frac{M^2}{2 \nu C}. 
\end{eqnarray*}
Using the Gronwall lemma (see {\sc Hirsch} and {\sc Smale}
\cite{smale:1}) leads to the differential inequality
\begin{eqnarray*}
\frac{d {\tilde K}({\tilde {\bv}})}{dt} &\le& \frac{M^2}{2 \nu C} e^{-
\nu C t},
\end{eqnarray*}
which, when solved, yields 
\begin{eqnarray*}
{\tilde K}({\tilde {\bv}}) &\le& e^{- \nu C t}{\tilde K}({\tilde {\bv}}
\mid_{t=t_0}) + \left( 1 - e^{- \nu C t} \right) \frac{M^2}{2 \nu^2
C^2}.
\end{eqnarray*}
This in turn implies
\begin{eqnarray*}
\lim_{t \rightarrow \infty} \sup {\tilde K}({\tilde {\bv}}) &\le&
\frac{M^2}{2 \nu^2 C^2}.
\end{eqnarray*}
The preceding analyses lead to natural notions of nonlinear dissipation
in the absence of forcing and long--term stability under conditions of
time--dependent loading for the analytic problem. 

\section{The Energetic Implications of the Time Discretisation}
\label{6}

This section is concerned with establishing a class of time
discretisations which inherit the self--same energetic properties
(nonlinear dissipation in the absence of forcing and long--term
stability under conditions of time dependent loading) as the analytic
problem, irrespective of the time increment employed. In this section a
generalised, Euler difference time--stepping scheme for the completely
general reference equation is formulated and the energetic
implications investigated in a similar vein as the analytic equations
in the previous section.

This stability analysis is inspired by the approach of others to
schemes for the conventional Navier--Stokes equations. The desirability
of the attributes identified as key energetic properties is recognised
and they have been used as a benchmark in the analysis of various of
the conventional, Eulerian Navier--Stokes schemes by a host of authors.
Related work on the conventional, Eulerian Navier--Stokes equations can
be found in a variety of references, for example {\sc Temam}
\cite{Temam:2} and {\sc Simo} and {\sc Armero} \cite{s:1}. 

The analyses presented here are extended, not only in the sense that
they deal with the completely general reference equation, but also in
that non--zero boundaries, so--called free boundaries and
time--dependent loads are able to be taken into account (the former two
as a consequence of the new inequality on page \pageref{94}). The
findings of this work have profound consequences for the implementation
of the deforming reference equations. It is significant that many
algorithms used for long--term simulation do not automatically inherit
the fundamental qualitative features of the dynamics.

\subsubsection*{A Generalised Time--Stepping Scheme}

An expression for a generalised Euler difference time--stepping scheme
can be formulated by introducing an ``intermediate'' velocity
\begin{eqnarray} \label{52}
{\tilde {\bv}}_{n + \alpha} \equiv \alpha \ {\tilde {\bv}}\mid_{t +
\Delta t} + (1 - \alpha) \ {\tilde {\bv}}\mid_{t} \hspace{5mm}
\mbox{for} \hspace{5mm} \alpha \in [0, 1]
\end{eqnarray}
to the variational momentum equation (equation (\ref{33}) on page
\pageref{33}) where ${\tilde {\bv}}\mid_t$ and ${\tilde {\bv}}\mid_{t +
\Delta t}$ are the solutions at times $t$ and $t + \Delta t$
respectively, $\Delta t$ being the time step. It is in this way that a
generalised time--discrete approximation of the momentum equation
\begin{eqnarray} \label{70}
&& \frac{{\rho}}{\Delta t} \left< {\tilde {\bw}}, ({\tilde {\bv}}_{n +
1} - {\tilde {\bv}}_{n }) {\tilde J}_{n + \alpha}
\right>_{L^2({\tilde \Omega}_{n + \alpha})} = \nonumber \\
&& \hspace{20mm} \left< {\tilde p} {\tilde \nabla} {\tilde {\bw}},
{\tilde {\bF}}_{n + \alpha}^{-t} {\tilde J}_{n + \alpha}
\right>_{L^2({\tilde \Omega}_{n + \alpha})} - 2 \mu \left< {\tilde
{\bD}}({\tilde {\bw}}), {\tilde {\bD}}({\tilde {\bv}}_{n + \alpha})
{\tilde J}_{n + \alpha} \right>_{L^2({\tilde \Omega}_{n +
\alpha})} \nonumber \\
&& \hspace{20mm} - {\rho} \left<{\tilde {\bw}}, ({\tilde \nabla}
{\tilde {\bv}}_{n + \alpha}) {\tilde {\bF}}_{n + \alpha}^{-1} \left(
{\tilde {\bv}}_{n + \alpha} - {\tilde {\bv}}_{n +
\alpha}^{\mbox{\scriptsize {\em ref}}} \right) {\tilde J}_{n + \alpha}
\right>_{L^2({\tilde \Omega}_{n + \alpha})} \nonumber \\
&& \hspace{20mm} + {\rho} \left<{\tilde {\bw}}, {\tilde {\bb}}_{n +
\alpha} {\tilde J}_{n + \alpha} \right>_{L^2({\tilde \Omega}_{n +
\alpha})} + \left< {\tilde{\bw}} , {\tilde{\bP}}_{n + \alpha} {\tilde
{\bN}_{n + \alpha}} \right>_{L^2({\tilde \Gamma}_{n + \alpha})}
\end{eqnarray}
is derived, where $\left< \ . \ \right>_{L^2({\tilde \Omega}_{n +
\alpha})}$ denotes the $L^2$ inner product over the deforming domain at
time $t + \alpha \Delta t$. ${\tilde \Gamma}_{n + \alpha}$, ${\tilde
{\bF}}_{n + \alpha}$, ${\tilde J}_{n + \alpha}$, ${\tilde {\bD}}_{n +
\alpha}$, ${\tilde {\bP}}_{n + \alpha}$, and ${\tilde {\bb}}_{n +
\alpha}$ are likewise defined to be the relevant quantities evaluated
at time $t + \alpha \Delta t$.

It will presently become clear that it makes sense to perform the
analyses for the time--discrete equation in the context of divergence
free rates of reference deformation only. This is since relevant
energy terms are not readilly recovered from the time-discrete
equations for deforming references in general. This investigation is
accordingly restricted to a subclass of reference deformations in
which ``reference volume'' is conserved. This is for reasons of
expedience alone and the subclass of deformations is thought to be
representative.

\begin{assumption} \label{assumption} The assumptions ${\tilde J}_{n} =
{\tilde J}_{n + \alpha}$ and ${\tilde J}_{n + 1} = {\tilde J}_{n +
\alpha}$ are made so that
\[
{\tilde K}({\tilde {\bv}}_n) = \frac{1}{2} \rho \left|\left| {\tilde
{\bv}}_n {\tilde J}_n^{\frac{1}{2}} \right|\right|_{L^2({\tilde
\Omega}_n)}^2 = \frac{1}{2} \rho \left|\left| {\tilde {\bv}}_{n}
{\tilde J}_{n + \alpha}^{\frac{1}{2}} \right|\right|_{L^2({\tilde
\Omega}_{n + \alpha})}^2 
\]
and
\[
{\tilde K}({\tilde {\bv}}_{n+1}) = \frac{1}{2} \rho \left|\left|
{\tilde {\bv}}_{n+1} {\tilde J}_{n+1}^{\frac{1}{2}}
\right|\right|_{L^2({\tilde \Omega}_{n+1})}^2 = \frac{1}{2} \rho
\left|\left| {\tilde {\bv}}_{n+1} {\tilde J}_{n +
\alpha}^{\frac{1}{2}} \right|\right|_{L^2({\tilde \Omega}_{n +
\alpha})}^2
\]
(by equation (\ref{52}) and since the volume of material over which
integration is being performed is constant). 
\end{assumption} 

{\sc Remark:} Notice that $\displaystyle \frac{{\tilde J}_{n + 1} -
{\tilde J}_{n}}{\Delta t} = {\tilde J} \mathop{\rm
div}{\bv}^{\mbox{\scriptsize {\it ref}}}_{n + \alpha}$, the discrete
form of $\displaystyle \displaystyle \frac{\partial {\tilde
J}}{\partial t} = {\tilde J} \mathop{\rm div}{\bv}^{\mbox{\scriptsize
{\it ref}}}$, can consequently be rewritten as 
\[
\mathop{\rm
div}{\bv}^{\mbox{\scriptsize {\it ref}}}_{n + \alpha} = 0
\]
under the conditions of the above assumption. It is for the practical
expedience afforded by Assumption \ref{assumption} alone that this
analysis is limited to instances in which $\mathop{\rm
div}{\bv}^{\mbox{\scriptsize {\it ref}}}_{n + \alpha} = 0$.

The following lemma will establish that the rate of energy change
associated with the convective term vanishes as a result of the
assumption.

\begin{lemma} [Discrete Convective Energy Rate] \label{354}
The discrete convective term 
\[
- {\rho} \left<{\tilde {\bw}}, ({\tilde \nabla}
{\tilde {\bv}}_{n + \alpha}) {\tilde {\bF}}_{n + \alpha}^{-1} \left(
{\tilde {\bv}}_{n + \alpha} - {\tilde {\bv}}_{n +
\alpha}^{\mbox{\scriptsize {\em ref}}} \right) {\tilde J}_{n + \alpha}
\right>_{L^2({\tilde \Omega}_{n + \alpha})}
\]
vanishes under circumstances of $\mathop{\rm
div}{\bv}^{\mbox{\scriptsize {\it ref}}}_{n + \alpha} = 0$ and a purely
Lagrangian description is used at free boundaries (alternatively
boundaries are of the fixed, impermeable type).
\end{lemma}
{\sc Proof:} The operator $\left< \ \cdot \ , ({\tilde \nabla} \ \cdot
\ ) {\tilde {\bF}}^{-1} {\tilde {\bw}} {\tilde J} \right>_{L^2({\tilde
\Omega})}$ is skew--symmetric for
\[
{\tilde {\bw}} \in W = \left\{ {\tilde{\bw}} \ : \ ({\tilde
\nabla} {\tilde{\bw}}) : {\tilde {\bF}}^{-t} = 0 \  \mbox{ on } {\tilde
\Omega}; \ {\tilde {\bw}} = {\bf 0} \mbox{ or } {\tilde
{\bF}}^{-t}{\tilde {\bN}} \cdot {\tilde {\bw}} = 0 \mbox{ on } {\tilde
\Gamma} \right\}
\]
by equation (\ref{309}) on page \pageref{309}. In instances where a
purely Lagrangian description is used to track free boundaries ${\tilde
{\bv}}_{n + \alpha} - {\tilde {\bv}}^{\mbox{\scriptsize {\em ref}}}_{n
+ \alpha}$ vanishes. At fixed, impermeable boundaries ${\tilde
{\bF}}_{n + \alpha}^{-t}{\tilde {\bN}}_{n + \alpha} \cdot {\tilde
{\bv}}_{n + \alpha}$ vanishes. The condition at the boundary is
therefore satisfied, under all of the afore--mentioned circumstances.
Apply the stipulated condition ${\tilde \nabla}
{\tilde{\bv}}^{\mbox{\scriptsize {\em ref}}} : {\tilde {\bF}}^{-t} = 0$
and set ${\tilde {\bw}} = {\tilde {\bv}}_{n + \alpha} - {\tilde
{\bv}}^{\mbox{\scriptsize {\it ref}}}_{n + \alpha}$ etc.

{\sc Remark:}\label{assumpremark} Recall that in the investigation of
the analytic problem, a term arising from the manipulation of the
acceleration containing term (the term containing the rate of change
of the Jacobian) cancelled with the convective energy. It is
therefore not surprising that assumptions pertaining to the
acceleration containing term (in particular to the rate of change of
the Jacobian) in the discrete problem will, once made, also be
necessary for the corresponding discrete convective energy term to
vanish (reffering to the $\mathop{\rm div}{\bv}^{\mbox{\scriptsize
{\it ref}}} = 0$ condition of Lemma \ref{354}). This is a good
prognosis for the energetic behaviour of the discrete problem in
circumstances of reference deformations excluded by Assumption
\ref{assumption}.

This concludes the preliminaries required for the analysis of the
time--discrete equation.

\subsection{Nonlinear Dissipation in the Absence of Forcing}

The following analysis establishes a class of time--stepping schemes
which exhibit nonlinear dissipation in the absence of forcing
regardless of the time increment employed.

\begin{theorem}[Nonlinear Dissipation in the Absence of Forcing]
\label{55} Suppose that the description is pure Lagrangian at any free
boundaries and that the deformation rate of the reference is divergence
free. A sufficient condition for the kinetic energy associated with the
generalised class of time--stepping schemes to decay nonlinearly
\begin{eqnarray*}
{\tilde K}({\tilde {\bv}}_{n + 1}) - {\tilde K}({\tilde {\bv}}_{n })
&\le& - {\Delta t} \ 2 \mu \left|\left| {\tilde {\bD}}({\tilde
{\bv}}_{n + \alpha}) {\tilde J}_{n + \alpha}^{\frac{1}{2}} 
\right|\right|_{L^2({\tilde \Omega}_{n + \alpha})}^2
\end{eqnarray*}
in the absence of forcing and irrespective of the time increment
employed, is that the scheme is as, or more, implicit than central
difference. That is
\[
\alpha \ge \frac{1}{2}.
\] 
\end{theorem}
{\sc Proof:} Expressing the intermediate velocities ${\tilde {\bv}}_{n
+ \frac{1}{2}}$ and ${\tilde {\bv}}_{n + \alpha}$ in terms of equation
(\ref{52}) and subtracting, the result
\begin{eqnarray} \label{48}
{\tilde {\bv}}_{n + \alpha} = \left( \alpha - \frac{1}{2} \right)
\left( {\tilde {\bv}}_{n  + 1} - {\tilde {\bv}}_{n } \right) +
{\tilde {\bv}}_{n + \frac{1}{2}}
\end{eqnarray}
is obtained. The first step towards formulating an expression
involving the kinetic energy of the generalised time stepping--scheme
(\ref{70}) is to replace the arbitrary vector, ${\bw}$, with ${\tilde
{\bv}}_{n + \alpha}$. By further substituting (\ref{48}) into
(\ref{70}) and eliminating the pressure containing term in a similar
manner to that in Theorem \ref{60}, an expression involving the
difference in kinetic energy over the duration of a single time step
is obtained.

The vector ${\tilde {\bv}} - {\tilde {\bv}}^{\mbox{\scriptsize {\em
ref}}}$ vanishes in instances where a purely Lagrangian description is
used to track free boundaries. The quantity ${\tilde {\bF}}_{n +
\alpha}^{-t}{\tilde {\bN}}_{n + \alpha} \cdot {\tilde {\bv}}_{n +
\alpha}$ vanishes where boundary conditions are of a fixed impermeable
type. The condition at the boundary for Lemma \ref{354} is therefore
satisfied. Incompressibility and a restriction on reference
deformations to those for which ${\mathop {\rm div}}{\bv}^{\scriptsize
ref}_{n+\alpha}$ is zero ensure that the remaining Lemma \ref{354}
condition is satisfied.

The equation 
\begin{eqnarray} \label{50}
{\tilde K}({\tilde {\bv}}_{n  + 1}) - {\tilde K}({\tilde {\bv}}_{n })
&=& - {\rho} \left( \alpha - \frac{1}{2} \right) \left|\left| \left(
{\tilde {\bv}}_{n  + 1} - {\tilde {\bv}}_{n} \right) {\tilde J}_{n +
\alpha}^{\frac{1}{2}} \right|\right|_{L^2({\tilde \Omega}_{n +
\alpha})}^2 \nonumber \\
&& - {\Delta t} \ 2 \mu \left|\left| {\tilde {\bD}}({\tilde {\bv}}_{n +
\alpha}) {\tilde J}_{n + \alpha}^{\frac{1}{2}} 
\right|\right|_{L^2({\tilde \Omega}_{n + \alpha})}^2 \nonumber +
{\Delta t} {\rho} \left< {\tilde {\bv}}_{n + \alpha}, {\tilde {\bb}}_{n
+ \alpha} {\tilde J}_{n + \alpha} \right>_{L^2({\tilde \Omega}_{n +
\alpha})} \nonumber \\
&& + {\Delta t} \left< {\tilde {\bv}}_{n + \alpha}, {\tilde{\bP}}_{n +
\alpha} {\tilde{\bN}}_{n + \alpha} \right>_{L^2({\tilde \Gamma}_{n +
\alpha})},
\end{eqnarray}
is then obtained. Since it is assumed that there is no forcing,
\begin{eqnarray*}
{\tilde K}({\tilde {\bv}}_{n + 1}) - {\tilde K}({\tilde {\bv}}_{n })
&\le& - {\rho} \left( \alpha - \frac{1}{2} \right) \left|\left| \left(
{\tilde {\bv}}_{n + 1} - {\tilde {\bv}}_{n} \right) {\tilde J}_{n +
\alpha}^{\frac{1}{2}} \right|\right|_{L^2({\tilde \Omega}_{n +
\alpha})}^2 \\
&& - {\Delta t} \ 2 \mu \left|\left| {\tilde {\bD}}({\tilde {\bv}}_{n +
\alpha}) {\tilde J}_{n + \alpha}^{\frac{1}{2}}
\right|\right|_{L^2({\tilde \Omega}_{n + \alpha})}^2.
\end{eqnarray*} 
Thus the kinetic energy inherent to the algorithmic
flow decreases nonlinearly in the absence of forcing, irrespective of
the time increment employed and for arbitrary initial conditions
provided that
\begin{eqnarray*}
\alpha \ge \frac{1}{2} \hspace{10mm} \mbox{and} \hspace{10mm}
\mathop{\rm div} {\bv}_{n + \alpha}^{\mbox{\scriptsize {\em ref}}} =
0.
\end{eqnarray*}
The former requirement translates directly into one specifying the use
of schemes as, or more, implicit than central difference. Only for
descriptions which become fully Lagrangian at free boundaries can it be
guaranteed that energy will not be artificially introduced by way of
the reference.

{\sc Remark:} Notice (by Lemma \ref{58}) that for $\alpha =
\frac{1}{2}$ an identical rate of energy decay
\begin{eqnarray*}
\frac{{\tilde K}({\tilde {\bv}}_{n + 1}) - {\tilde K}({\tilde {\bv}}_{n
})}{{\Delta t}} &\le& - 2 \nu C {\tilde K}({\tilde {\bv}}_{n + \alpha})
\end{eqnarray*} 
is obtained for the discrete approximation as was obtained for the
equations.

\subsection{Long--Term Stability under Conditions of Time--Dependent
Loading}

This second part of the time--discrete analysis establishes a class of
time stepping schemes which exhibit long--term stability under
conditions of time dependent loading irrespective of the time increment
employed. The following lemma is necessary to the analysis and is
concerned with devising a bound for the energy at an intermediate point
in terms of energy values at either end of the time step.
\begin{lemma}[Intermediate Point Energy] \label{57} The following bound
applies
\[
{\tilde K}({\tilde {\bv}}_{n + \alpha}) \ge \alpha \left( \alpha - c +
\alpha c \right) {\tilde K}({\tilde {\bv}}_{n  + 1}) + (1-\alpha)
\left( 1-\alpha - \displaystyle \frac{\alpha}{c} \right) {\tilde
K}({\tilde {\bv}}_{n })
\]
where $c$ is some constant, $c > 0$.
\end{lemma}
{\sc Proof:} By Young's inequality
\begin{eqnarray} \label{51}
\left|\left| {\tilde {\bv}}_{n+1} {\tilde J}_{n +
\alpha}^{\frac{1}{2}}  \right|\right|_{L^2({\tilde \Omega}_{n +
\alpha})} \ \left|\left| {\tilde {\bv}}_{n} {\tilde J}_{n +
\alpha}^{\frac{1}{2}}  \right|\right|_{L^2({\tilde \Omega}_{n +
\alpha})} &\le& \left(\frac{c}{2}\right) \left|\left| {\tilde
{\bv}}_{n+1} {\tilde J}_{n + \alpha}^{\frac{1}{2}}
\right|\right|_{L^2({\tilde \Omega}_{n + \alpha})}^2 \nonumber \\
&& + \left( \frac{1}{2c} \right) \left|\left| {\tilde {\bv}}_{n}
{\tilde J}_{n + \alpha}^{\frac{1}{2}}  \right|\right|_{L^2({\tilde
\Omega}_{n + \alpha})}^2
\end{eqnarray}
for $c > 0$. Writing ${\tilde K}({\tilde {\bv}}_{n+\alpha})$
explicitly in terms of the ``intermediate'' velocity definition,
(\ref{52}), leads to
\begin{eqnarray*}
{\tilde K}({\tilde {\bv}}_{n + \alpha}) &=& \alpha^2 {\tilde K}({\tilde
{\bv}}_{n  + 1}) + (1 - \alpha)^2 {\tilde K}({\tilde {\bv}}_{n }) + 2
\alpha(1-\alpha) \left< {\tilde {\bv}}_{n  + 1},{\tilde {\bv}}_{n}
{\tilde J}_{n + \alpha} \right>_{L^2({\tilde \Omega}_{n + \alpha})} \\
&\ge& \alpha^2 {\tilde K}({\tilde {\bv}}_{n + 1}) + (1 - \alpha)^2
{\tilde K}({\tilde {\bv}}_{n }) \\
&& - 2 \alpha(1-\alpha)\left|\left| {\tilde {\bv}}_{n + 1} {\tilde
J}_{n + \alpha}^{\frac{1}{2}} \right|\right|_{L^2({\tilde \Omega}_{n +
\alpha})} \ \left|\left| {\tilde {\bv}}_{n} {\tilde J}_{n +
\alpha}^{\frac{1}{2}}  \right|\right|_{L^2({\tilde \Omega}_{n +
\alpha})} \\
&\ge& \alpha \left[ \alpha - (1 - \alpha) c \right] {\tilde K}({\tilde
{\bv}}_{n  + 1}) + (1-\alpha) \left[ (1-\alpha) - \frac{\alpha}{c}
\right] {\tilde K}({\tilde {\bv}}_{n })
\end{eqnarray*}
using equation (\ref{51}). The optimal choice of the constant $c$ is
established farther on.

The following theorem establishes a class of time--stepping schemes
which exhibit long--term stability under conditions of time--dependent
loading regardless of the time increment employed.

\begin{theorem}[Long--Term Stability] \label{155}
Suppose that the description is pure Lagrangian at any free boundaries
and that the rate at which the reference is deformed is divergence
free. A sufficient condition for the algorithmic flow to exhibit
long--term stability under conditions of time--dependent loading
(intrinsic to real flows and the Navier--Stokes equations), assuming
this time--dependent loading and the speed of the free surface is
bounded in such a way that
\begin{eqnarray*}
{\rho} \left|\left| {\tilde {\bb}}_{n + \alpha} {\tilde J}_{n +
\alpha}^{\frac{1}{2}} \right|\right|_{L^2({\tilde \Omega}_{n +
\alpha})}^2 + \left|\left| {\tilde {\bP}}_{n + \alpha}{\tilde {\bN}}_{n
+ \alpha} \right|\right|_{L^2({\tilde \Gamma}_{n + \alpha})}^2 + \nu^2
C^2 \left|\left| {\tilde {\bv}}_{n + \alpha}
\right|\right|_{L^2({\tilde \Gamma}_{n + \alpha})}^2 &\le& M^2,
\end{eqnarray*}
is
\[
\alpha > \frac{1}{2}.
\] 
\end{theorem}
{\sc Proof:} Substituting Lemma \ref{49} (page \pageref{49}) and Lemma
\ref{58} (page \pageref{58}) into equation (\ref{50}), applying the
above bound and choosing $\alpha \ge \frac{1}{2}$ one obtains
\begin{eqnarray*}
\frac{{\tilde K}({\tilde {\bv}}_{n  + 1}) - {\tilde K}({\tilde
{\bv}}_{n })}{\Delta t} + \nu C {\tilde K}({\tilde {\bv}}_{n + \alpha})
&\le& \frac{M^2}{2 \nu C}.
\end{eqnarray*}
From this point on the argument used is identical to that of {\sc
Simo} and {\sc Armero} \cite{s:1} for the conventional, Eulerian
Navier--Stokes equations.  Substitution of Lemma \ref{57} leads to a
recurrence relation,
\begin{eqnarray*}
{\tilde K}({\tilde {\bv}}_{n + 1}) &\le& \frac{1 - \nu C (1 - \alpha)(1
- \alpha - \frac{\alpha}{c})\Delta t}{1 + \nu C\alpha (\alpha - c +
\alpha c)\Delta t}{\tilde K}({\tilde {\bv}}_{n }) + \frac{M^2 \Delta
t}{2 \nu C \left[ 1 + \nu C\alpha (\alpha - 1 + \alpha c)\Delta t
\right]}.
\end{eqnarray*}
Using this recurrence relation to take cognisance of the energy over
all time steps,
\begin{eqnarray} \label{200}
{\tilde K}({\tilde {\bv}}_{n + 1}) &\le& \left[ \frac{1 - \nu C(1 -
\alpha)(1 - \alpha - \frac{\alpha}{c})\Delta t}{1 + \nu C\alpha (\alpha
- c + \alpha c)\Delta t} \right]^{n} {\tilde K}({\tilde {\bv}}_0)
\nonumber \\
&& + \frac{M^2 \Delta t}{2\nu C \left[ 1 + \nu C\alpha (\alpha - c +
\alpha c)\Delta t \right]} \sum_{k=0}^{n-1} \left[ \frac{(1 - \nu C(1 -
\alpha)(1 - \alpha - \frac{\alpha}{c})\Delta t)}{1 + \nu C\alpha
(\alpha - c + \alpha c)\Delta t} \right]^{k} \nonumber \\
&& \hspace{133mm}
\end{eqnarray}
is obtained. An infinite geometric series which converges so that
\begin{eqnarray*}
\lim_{n \rightarrow \infty} \sup {\tilde K}({\tilde {\bv}}_{n  + 1})
&\le& \frac{M^2 \Delta t}{2 \nu C \left[ 1 + \nu C \alpha (\alpha - c +
\alpha c)\Delta t \right]} \left[ 1 \frac{}{}^{}_{} \right. \\
&& \hspace{30mm} \left. - \ \frac{(1 - \nu C(1 - \alpha)(1 - \alpha -
\frac{\alpha}{c})\Delta t)}{1 + \nu C \alpha (\alpha - c + \alpha
c) \Delta t} \right]^{-1} \\
&& \\
&=& \frac{M^2}{2\nu C \left[ \nu C\alpha (\alpha - c + \alpha c) + \nu
C(1 - \alpha)(1 - \alpha - \frac{\alpha}{c}) \right]}
\end{eqnarray*}
results, providing the absolute ratio of the series is less than
unity. That is
\[
\left| \frac{1 - \nu C(1 - \alpha)(1 - \alpha - \frac{\alpha}{c})\Delta
t}{1 + \nu C\alpha (\alpha - c + \alpha c)\Delta t} \right| < 1.
\]
Therefore either
\[
- 1 - \nu C\alpha(\alpha - c + \alpha c)\Delta t < 1 - \nu C(1 -
\alpha) \left (1 - \alpha - \frac{\alpha}{c} \right) \Delta t
\]
or
\begin{eqnarray} \label{202}
1 - \nu C(1 - \alpha) \left (1 - \alpha - \frac{\alpha}{c} \right)
\Delta t < 1 + \nu C\alpha(\alpha - c + \alpha c)\Delta t
\end{eqnarray}
in order for the bound to exist. Notice, furthermore, that for this
desired convergence to be unconditional (regardless of the time
increment employed) requires
\begin{eqnarray} \label{201}
\alpha - c + \alpha c \ge 0.
\end{eqnarray} 
The denominator in the series ratio might otherwise vanish for some
value of $\Delta t$.

For $\alpha \in \left[\frac{1}{2}, 1\right]$ equation (\ref{202}) and
equation (\ref{201}) together imply
\begin{eqnarray*}
\frac{(1 - \alpha)}{\alpha} < c \le \frac{\alpha}{(1 -
\alpha)} 
\end{eqnarray*}
which in its turn implies
\begin{eqnarray*}
\frac{(1 - \alpha)}{\alpha} < \frac{\alpha}{(1 - \alpha)}.
\end{eqnarray*}
The choice of the parameter $\alpha > \frac{1}{2}$ therefore leads to
an infinite geometric series which forms the desired upper bound. The
minimum value of this bound occurs for $c$ chosen according to
\begin{eqnarray*}
\inf_{\frac{(1-\alpha)}{\alpha} < c \le \frac{\alpha}{(1-\alpha)}}
\frac{1}{\nu C\alpha(\alpha - c + \alpha c) + \nu C(1 - \alpha)(1 -
\alpha - \frac{\alpha}{c})} &=& \frac{1}{\nu C (2\alpha - 1)^2}.
\end{eqnarray*}
The value of this upper bound, which occurs for the choice of the
parameter $\alpha > \frac{1}{2}$, is then
\begin{eqnarray*}
\lim_{n \rightarrow \infty} \sup {\tilde K}({\tilde {\bv}}_{n  + 1})
&\le& \frac{M^2}{2 \nu^2 C^2 (2\alpha - 1)^2}.
\end{eqnarray*}
In this way one arrives at a class of algorithms which are
unconditionally (irrespective of the time increment employed) stable.

{\sc Remark:} Notice that for $\alpha = 1$ one obtains an identical
energy bound for the discrete approximation as was obtained for the
equations.

\section{An ``Updated'' Approach and a Simplified Implementation}
\label{44}

Ever burgeoning deformation gradients accumulate for a straight forward
implementation of the equations. Using an ``updated'' approach is one
way of coping with this otherwise rather daunting prospect. An
``updated'' approach is the result of a little, well--worthwhile
lateral thinking. An ``updated'' approach amounts to choosing a new
referential configuration during each time step.

In the case of time stepping schemes based about a single instant (eg.
the generalised class of Euler difference schemes investigated in
Section \ref{6}) a considerably simplified implementation can further
be achieved by a particularly appropriate choice of configurations.
Making the choice of a referential configuration which coincides with
the spatial configuration at the instant about which the time stepping
scheme is based allows the deformation gradient to be omitted
altogether (the deformation gradient is identity under such
circumstances). For such implementations (those which require
evaluation about a single point only) no error arises from the use of
the equations cited in {\sc Hughes, Liu} and {\sc Zimmerman}
\cite{h:1},
\begin{eqnarray} 
{\rho} \left( \frac{\partial { {\bv} }}{\partial t} + {{{
\nabla}  {\bv} }  } ({ {\bv} } - { {\bv} }^{ref})
\right) &=& {\rho} { {\bb} } + \mathop{ {\rm
div}}{ {\bsigma}} \label{307} \\
\mathop{ {\rm div}}{ {\bv}} &=& 0. \label{308}
\end{eqnarray}
These equations are not valid for any, arbitrary choice of reference
or if the implementation requires the equation to be evaluated at
more than one point within each time step (eg. a Runge--Kutta or
finite--element--in--time scheme). It is important to remember that
in a discrete context the reference configuration is fixed for the
duration of the entire time increment. Although the referential
configuration is hypothetical and can be chosen arbitrarily for each
time step, once chosen it is static for the duration of the entire
time step. Once the coincidence of configurations is ordained at a
given instant, ${\tilde {\bF}}$ is defined by the deformation, both
before and after, and must be consistant.

There would seem to be no reason why one would wish to define the
deformation about a configuration other than that at the instant about
which the implementation is based (assuming the implementation used is
indeed based about a single point eg. a finite difference) thereby
involving deformation gradients. Resolving the resulting difficulties
associated with the deformation gradients by means of a perturbation
seems unnecessarily complicated in the light of the above reasoning.

\section{Conclusions}

The correct equations, which describe the motion of an
incompressible, Newtonian fluid and which are valid for a completely
general range of reference deformations, are equations (\ref{29}) and
(\ref{30}). For implementations requiring the equations to be
evaluated about a single instant within each time step only (eg.
finite differences), the deformation gradients may be assumed
identity i.e. the equations of {\sc Hughes, Liu} and {\sc Zimmerman}
\cite{h:1} (equations (\ref{307}) and (\ref{308})) will suffice.

In this work it is shown (as was hoped) that nonlinear,
exponential--type dissipation in the absence of forcing and long--term
stability under conditions of time dependent loading are properties
automatically inherited by deforming reference descriptions. The single
provisor is that such descriptions become fully Lagrangian at any
moving boundaries. These properties are intrinsic to real flows and the
conventional, Eulerian Navier--Stokes equations. 

Relevant energy terms are not readily recovered from the
time--discrete equations for deforming references in general. Only for
divergence free rates of reference deformation which become fully
Lagrangian at free boundaries could it consequently be guaranteed that
energy would not be artificially introduced to the algorithmic flow by
way of the reference. The divergence free assumption was made for
reasons of expedience alone and the limitations of the time--discrete
analysis are consequently not expected to detract from the use of the
method in any way. This is especially so when it is considered that, a
term arising from the manipulation of the acceleration containing term
(the term containing the rate of change of the Jacobian) cancelled with
the convective energy in the investigation of the analytic problem and
that assumptions pertaining to the acceleration containing term (in
particular to the rate of change of the Jacobian) in the discrete
problem were, once made, also necessary for the corresponding discrete
convective energy term to vanish (reffering to the $\mathop{\rm
div}{\bv}^{\mbox{\scriptsize {\it ref}}} = 0$ condition of Lemma
\ref{354}). If one were to be overly cautious on this basis one would
be faced with the additional challenge of enforcing a fully Lagrangian
description for nodes situated on any free boundaries, while deforming
elements would be required to deform at a rate which is divergence
free. Such a totally divergence free description may, however, not be
possible. An alternative strategy would be to use a fully Lagrangian
description. Both the purely Lagrangian and purely Eulerian fluid
descriptions have divergence free rates of distortion.

There are inherent problems with using certain classes of
time--stepping schemes and the use of finite difference schemes more
implicit than central difference is consequently advocated. Such
differences exhibit the key energetic properties (nonlinear,
exponential--type dissipation in the absence of forcing and long--term
stability under conditions of time dependent loading) irrespective of
the time increment employed. A backward difference is the obvious
choice. Calculations at time $t + \alpha \Delta t$ would require an
intermediate mesh and associated quantities for instances in which
$\alpha \ne 1$ (since $\alpha > \frac{1}{2}$).

The author recommends a strategy in which a predominantly Eulerian
description is used, where possible, for the bulk of the problem (from
an efficiency point of view) and the completely general reference
description for the remainder is appropriate. Purely Eulerian
descriptions have the advantage of a ``one off'' finite element
construction and involve none of the hazards of a badly distorted
reference.

\section{Acknowledgements}

Grzegorz Lubczonok and Ronald Becker are thanked for their respective opinions on the inequality on page \pageref{94}, as is Daya Reddy. The use of Kevin Colville/George Ellis' printer is also gratefully acknowledged.

\bibliography{cmame1}

\end{document}